%% file: conj.tex
\documentclass[12pt,a4paper]{article}

\include{packlist}
\include{defs}

\newcommand\bref[1]{(\ref{#1})}
\numberwithin{equation}{section}
\usepackage{authblk}
\author[a,b]{Ruth Britto}
\author[a]{Guy~R.~Jehu}
\author[a]{Andrea Orta}
\affil[a]{\textit{School of Mathematics and Hamilton Mathematical 
Institute,}}
\affil[ ]{\textit{Trinity College Dublin, 
Dublin 2, Ireland}}
\affil[b]{\textit{Institut de Physique Th\'eorique,
		Universit\'e Paris Saclay,
		CEA, CNRS, F-91191 Gif-sur-Yvette
		cedex, France
}}
\title{The dimension-shift conjecture for
 one-loop amplitudes}
\date{\today}
\begin{document}
\maketitle
\setcounter{page}{2}
\pagenumbering{arabic}
\begin{abstract}
	A conjecture made by Bern, Dixon, Dunbar, 
	and Kosower asserts a 
	simple dimension shifting relationship between the 
	one-loop structure 
	of $\mathcal{N}=4$ MHV amplitudes
	and all-plus helicity amplitudes in 
	pure Yang-Mills theory. 
	We prove this conjecture to all orders in dimensional
	regularisation using unitarity cuts, 
	and evaluate the
	form of these simplest one-loop amplitudes using
	a generalised $D$-dimensional unitarity
	technique which captures the full amplitude 
	to all multiplicities.
\end{abstract}

\input{intro-sec.tex}


\input{proof-sec.tex}

\input{cuts-sec.tex}

\input{conclusion-sec.tex}

\section*{Acknowledgments}

We would like to thank Simon Badger, David Dunbar, Riccardo Gonzo, Martijn Hidding, Robert Schabinger, and Johann Usovitsch for
 many helpful discussions. 
Diagrams were drawn using Axodraw.
This work was supported by the European Research Council through grant number 647356 (CutLoops).

\bibliographystyle{nb.bst}
\bibliography{biblo}{}

\end{document}

%% file: packlist.tex
\usepackage{amsmath,amssymb}
\usepackage{setspace}
\usepackage{axodraw}
\usepackage{slashed}
\usepackage{hyperref}
\usepackage{bm}
\usepackage{mathtools}
\usepackage[margin=1in]{geometry}
\usepackage{titlesec}
\usepackage{xcolor}
\usepackage{cite}
\titleformat{\section}
  {\normalfont\Large\bfseries}{\thesection}{1em}{}[{\color{gray}\titlerule[0.4pt]}]

%% file: defs.tex
\def\la{\langle}
\def\ra{\rangle}
\def\spa#1.#2{\left\langle#1\,#2\right\rangle}
\def\spb#1.#2{\left[#1\,#2\right]}
\def\spba#1.#2.#3{\left[#1|#2|#3\right\rangle}

\newcommand{\mi}{\scalebox{0.94}[1.0]{-}}

\newcommand{\tree}{\text{tree}}

\def\gg{p_\chi}

\DeclareMathOperator{\tr}{\mathrm{tr}}

\def\notag{\nonumber}



%% file: intro-sec.tex

\section{Introduction}
Planar $\mathcal{N}=4$ super Yang-Mills continues to prove itself  
as the simplest toy model to develop techniques for computing 
scattering amplitudes from their singular structure; 
recent work has pushed analytic
ans\"{a}tze for planar amplitudes up to seven points at 
four loops~\cite{Caron-Huot:2019bsq,Dixon:2020cnr}.
A remarkable feature which entails this particular 
facet of its simplicity
is its on-shell constructibility~\cite{Elvang:2015rqa,Arkani-Hamed:2016byb}: 
the use of (four-dimensional) unitarity cuts to construct 
amplitudes~\cite{Bern:1994ju,
Bern:1994cg,Britto:2004nc,CaronHuot:2012ab}
finds its apotheosis in this theory, and this has 
led to many fruitful insights, but there are 
still outstanding questions about what broader 
statements about scattering amplitudes in 
more general quantum field theories 
can be induced from these successes.

Although QCD amplitudes remain a few steps behind in comparison,
a firm foothold has been gained at two-loop 
order for five-gluon 
scattering~\cite{Badger:2013gxa,Badger:2017jhb,Abreu:2017hqn}
thanks to finite field numerical 
reconstruction of 
numerators~\cite{Peraro:2016wsq}
on $D$-dimensional unitarity cuts;
these results have in turn been processed into more manageable
functions~\cite{Badger:2018enw,Abreu:2018zmy,Abreu:2019odu}. 
A particularly manageable example at two loops 
is the all-plus helicity amplitude (AP amplitude for short). 
Due to its simpler analytic structure, particularly
when arranged to separate off the infrared (IR) divergence~\cite{Gehrmann:2015bfy}, 
the amplitude can in fact be
directly computed with one-loop 
methods~\cite{Dunbar:2016aux,Dunbar:2017nfy}.
Recent work on its subleading-colour 
structure~\cite{Badger:2019djh,Dunbar:2019fcq} has led 
to proofs of conformal invariance
at one loop~\cite{Henn:2019mvc},
and an all-multiplicity (all-$n$) 
form of a partial amplitude~\cite{Dunbar:2020wdh},
the first such example.

In 1996, Bern, Dixon, Dunbar and Kosower
(BDDK)~\cite{Bern:1996ja} 
conjectured a relation between the simplest one-loop gluon amplitudes in these two theories. 
When computed in $D=4-2\epsilon$ dimensions in dimensional regularisation, they conjectured that for arbitrary multiplicity $n$,
\begin{align}
	A^{{\rm QCD}}_n\left(1^+,2^+,...,n^+\right) =
	-2\epsilon(1-\epsilon)(4\pi)^2\left[
	\frac{A_n^{\mathcal{N}=4}
	(1^+,...,i^-,..,j^-,...,n^+)}{
	\la ij \ra^4}\right]_{\epsilon\rightarrow \epsilon-2}
	\; ,
	\label{eq:conj}
\end{align}
where  
the change in $\epsilon$ on the right-hand-side
corresponds to a ``dimension
shift" $4-2\epsilon\rightarrow 8-2\epsilon$. 
While the amplitudes on each side were by then well known up to $\mathcal{O}(\epsilon)$, there has been little motivation to develop methods to probe higher orders in $\epsilon$, and so it has been difficult to test the validity of the conjecture. 
A further motivation to do so comes from recent interest in properties of amplitudes viewed in terms of their expansions in multiple polylogarithms. It would be interesting to know whether these most accessible amplitudes, which are highly constrained by their myriad symmetries, are well behaved and easily predicted when probed at higher orders in $\epsilon$.

Both the all-plus and the MHV amplitudes  have close connections with integrability 
and string theory.
At one loop, the AP amplitude has been explicitly shown to be equivalent to 
the one-loop amplitude in self-dual Yang-Mills theory:
the latter was built from the Lagrangian by 
Cangemi~\cite{Cangemi:1996pf}, as well as by Chalmers and Siegel~\cite{Chalmers:1996rq}
which built on previous work by Bardeen~\cite{Bardeen:1995gk}, 
who computed four and five-point amplitudes directly from
all-plus tree level amplitudes with two off-shell legs, 
which is very close to the analysis we carry out in 
section~\ref{sec:cuts} to all multiplicity.

Moreover Bardeen proposed that the non-vanishing of the 
AP amplitude at loop level
was rooted in an anomaly associated with 
the currents in the self-dual sector;
recent work along these lines explores these 
elemental notions in a manner 
complementary to the ``amplitudes"-based techniques presented 
here~\cite{Chattopadhyay2020}.
Self-dual Yang-Mills has historically been 
a theory of interest as it is
classically integrable (see~\cite{Popov:1998pc} and references therein), and can 
also be obtained from the compactification of
$\mathcal{N}=2$ strings in $2+2$ spacetime 
dimensions~\cite{Ooguri1995}.

The technique used by BDDK in~\cite{Bern:1996ja} for 
computing the all-epsilon structure
of $\mathcal{N}=4$ MHV amplitude for $n=5,6$ built upon the computation
of the four-point string amplitude by Green, Schwarz and 
Brink~\cite{Green:1982sw}, by taking the field-theory
limit of the 10-dimensional open superstring 
to compute the amplitude in $D=4-2\epsilon$.
Systematic rules which generalise a string-based computation
of field-theory amplitudes were developed 
by Bern and Kosower~\cite{Bern:1991aq}, and it was shown that
they can also be directly derived from
the Schwinger worldline formalism as derived by Bern and 
Dunbar~\cite{Bern:1991an}.
As a computational alternative to Feynman diagrams worldline
approaches
maintain significant 
interest~\cite{Ahmadiniaz2020,Ahmadiniaz2020a}
providing a context to study asymptotic states
in scattering problems~\cite{Bonocore2021},
and for computing gravitational-radiation contributions 
to black-hole scattering~\cite{Mogull2020}.

Statements concerning the relationship 
between $\mathcal{N}=4$ MHV and 
AP amplitudes can also be found in the literature.
Schabinger~\cite{Schabinger2011} has pointed out that 
dimension-shifted one-loop $\mathcal{N}=4$ amplitudes
occur in the tree-level $\mathcal{O}(\alpha'^2)$ string computations
of Stieberger and Taylor~\cite{Stieberger2006}, and that 
as both amplitudes are dominated by cyclic structure and $F^4$ 
terms, they are so constrained that they must coincide. 
This provides a strong argument for the validity of 
equation~\bref{eq:conj}, and indeed the arguments we make in 
section~\ref{sec:cuts} confirm this.

\subsection{Overview}
In supersymmetric Yang-Mills theories,
the all-plus (and indeed the single-minus) gluon amplitudes can be shown to
vanish to all
orders in perturbation theory\footnote{As the tree level amplitudes coincide
in SUSY and in QCD, this is also a statement about massless QCD at leading order.}
thanks to supersymmetric (SUSY) Ward identities~\cite{Grisaru:1977px},
\begin{align}
	A_n^{\rm SUSY}(1^\pm,2^+,...,n^+) = 0 \quad .
	\label{eq:apvan}
\end{align}

At one-loop order we can decompose the  gluonic ($[1]$), fermionic ($[1/2]$) 
contributions to the loop content into 
terms which can be expressed entirely in terms of amplitudes in supersymmetric theories,
and an independent ($\mathcal{N}=0$) part which corresponds to the contribution from a 
complex scalar~\cite{Dixon1996}
\begin{align}
	A^{[0]} &= \mathcal{S} \; , 
	\\
	A^{[{1\over 2}]} &= -2\mathcal{S} + \mathcal{F} \; ,
	\\
	A^{[1]} & = 2\mathcal{S} +\mathcal{G} \quad .
	\label{eq:sdec}
\end{align}
Here $A^{[0]}$ denotes the contribution from a real scalar: as the 
scalar and anti-scalar contributions are equal, the complex case 
is retrieved by simply doubling this term.
The expressions~\bref{eq:sdec} can be seen emerge from inverting the 
decompostion of the supersymmetric 
multiplets\footnote{They also emerge naturally from 
implementing the Bern-Kosower rules in the string-based 
formalism~~\cite{Bern:1991aq,Bern:1991an,Bern1992a}.}
\begin{align}
	A^{\mathcal{N}=4} &=A^{[1]} + 4A^{[{1\over 2}]} 
			+ 6A^{[0]}
			\label{eq:pdec}
			\\
			&=\mathcal{G}+ 4\mathcal{F}
			\label{eq:Ne4}
			\\
	A^{\mathcal{N}=1} &=A^{[1]} + A^{[{1\over 2}]} 
			\notag \\
			&=\mathcal{G}+ \mathcal{F}\quad .
\end{align}
From the vanishing property~\bref{eq:apvan} we conclude that
\begin{align}
	A_n^{[1]}(++...+) &=2A^{[0]} =2\mathcal{S} \; .
	\label{eq:apscal}
\end{align}
A major simplification is thus made manifest: the AP
amplitude is equivalent to having just two real
scalars circulating around the loop.
This makes its computation using 
massive cuts particularly tractable~\cite{Badger:2008cm}.
It is thus puzzling that in this context the $\mathcal{N}=4$ MHV 
amplitude seems to have a structure almost entirely converse to the 
all-plus in that equations~\bref{eq:apscal} and~\bref{eq:Ne4} 
have mutually disjoint content. 

The conjecture~\bref{eq:conj} was verified by BDDK up to six points, 
with the explicit expressions for the QCD amplitudes computed to be
\begin{align}
	A^{\rm AP}_4 \equiv & A_4^{\rm QCD}\left(1^+,2^+,3^+,4^+\right)= {-2\epsilon(1-\epsilon)\over
	\la 12\cdots 41\ra} s_{12}s_{23} I^{8-2\epsilon}_4 \;,
	\label{eq:a4}
	\\
	A^{\rm AP}_5 =& {-\epsilon(1-\epsilon)\over
	2\la 12 \cdots 51 \ra}\left[
		\sum_{j=1}^n
		s_{j+1,j+2}s_{j+2,j+3}I_4^{8-2\epsilon,(j)} + (D-4)\tr_5(1234)I^{10-2\epsilon}_5\right]\; ,
	\label{eq:a5}
	\\
	A^{\rm AP}_6 =& {-\epsilon(1-\epsilon)\over
	2\la 12\cdots 61 \ra}\biggl[
		-\sum_{1<j_1<j_2\leq n}^n
		\tr\bigl((j_1+1)q_{j_1+1,j_2+1}(j_2+1)q_{j_2+1,j_1+1}\bigr)I_4^{8-2\epsilon,(j_1,j_2)} 
	\notag \\
	&	+ (4-2\epsilon)\left(\sum_{j=1}^n\tr_5(j+1,j+2,j+3,j+4)I^{10-2\epsilon,(j)}_5
	+\tr(123456)I^{10-2\epsilon}_6\right)\biggr]\; ,
\label{eq:a6}
\end{align}
and the $\mathcal{N}=4$ MHV amplitudes can be obtained
from~\bref{eq:conj}. Here we use the variables $q_{rs}$ to denote sums of consecutive momenta
$q_{rs}=\sum_{i=r}^{s-1}p_i$. 
The descendant integral notation $(j_1,j_2)$ denotes the shrinking of the propagators 
leading into the $j_1$th and $j_2$th leg of the maximal $n$-gon.
We use the usual notation for two-particle Mandelstam invariants $s_{ij} =(p_i+p_j)^2$, and
\begin{align}
	\tr(ab\cdots) &= \tr(\slashed{p}_a\slashed{p}_b\cdots)
		&= \tr_+(ab\cdots) +\tr_-(ab\cdots) \; ,
		\\
	\tr_5(ab\cdots) &= \tr(\gamma_5\slashed{p}_a\slashed{p}_b\cdots)
		&= \tr_+(ab\cdots) - \tr_-(ab\cdots) \; ,
\end{align}
where $\tr_\pm$ involve the usual chiral projectors, and 
can be expressed  in terms of spinor-helicity  
if one of the momenta is null. For example, if $p_a^2=0$ then
\begin{align}
	\tr_+(ab\cdots) &= {1\over 2}
		\tr((1+\gamma_5)\slashed{p}_a\slashed{p}_b\cdots)
		= [a|b\cdots|a\ra \; ,
		\notag \\
	\tr_-(ab\cdots) &= {1\over 2}
		\tr((1-\gamma_5)\slashed{p}_a\slashed{p}_b\cdots)
		= \la a|b\cdots|a]\quad .
\end{align}

Although 
the study of scattering amplitudes has advanced significantly since this
conjecture was first proposed, an explicit proof 
has not been presented until now.
The 
all-multiplicity expressions for both the MHV in $\mathcal{N}=4$  
and the AP amplitude were already known to leading order in $\epsilon$ at 
the time of the conjecture~\cite{Bern:1996ja}.
We review these results in a more 
contemporary framework.

If one is content with truncating terms $\sim\mathcal{O}(\epsilon)$, 
a one-loop amplitude that is 
dimensionally regulated in $4-2\epsilon$ dimensions
can be expressed in terms of  a basis of box, 
triangle and bubble integrals with algebraic functions as coefficients, 
and additional purely rational terms,
\begin{align}
	A_n = \mathbf{d}_4\cdot \mathbf{I}^{D=4-2\epsilon}_4 
	+ \mathbf{d}_3\cdot\mathbf{I}^{D=4-2\epsilon}_3
	+\mathbf{d}_2\cdot\mathbf{I}^{D=4-2\epsilon}_2
	+d_R +\mathcal{O}(\epsilon)\; ,
	\label{eq:bas1i}
\end{align}
where $\mathbf{d}_m\cdot\mathbf{I}_m$ are dot products weighting
the $m$-point scalar integrals, $I^{[i_1,...,i_m]}_m[1]$ with
algebraic functions of kinematic variables. 
The integral functions are defined as
\begin{align}
	I^{D;[i_1,...,i_m]}_m[1] &=i(-1)^{n+1}(4\pi)^{2-\epsilon}
	e^{\gamma_E\epsilon}\int{d^D\ell\over(2\pi)^D}
	{1\over \ell^2(\ell-q_{i_1i_2})^2
	\cdots (\ell-q_{i_1i_m})^2} \; .
    \label{eq:masint}
\end{align}
In general we denote the propagator just before the $k$th 
leg $\ell_{i_k}$ so in equation \bref{eq:masint} $\ell \equiv \ell_{i_1}$.
The normalisation is set such that after the loop integration
(but before the Feynman parameter integration) shift 
and reduction
identities~\cite{Bern:1993kr,Jehu:2020xip} are simplified,
consistent with~\cite{Abreu:2017ptx}. 
We suppress the term $+i0$ in each propagator.
From here on the absence of a $D$ superscript on $I_m$ implies
$D=4-2\epsilon$.
The argument in the square brackets, 
indicates possible numerator terms introduced:
these could be loop momenta or Feynman parameters and thus apply 
at different stages of the loop integration depending on context;
if no brackets are present the numerator is understood to be 1
and we speak of scalar integrals, as in the formula above.

The term $d_R$ in the basis~\bref{eq:bas1i} is a rational function of 
spinor-brackets and momentum variables only.
One manifestation of the simplicity of  amplitudes in $\mathcal{N}=4$ SYM
at one-loop  is the fact that
\begin{align}
	\mathbf{d}^{\mathcal{N}=4}_i = \mathbf{0}, 
	\quad {\rm for\;} i\in \lbrace R,2,3 \rbrace\; ;
	\label{eq:n4fact}
\end{align}
while for the all-plus amplitude, 
the lack of four-dimensional cuts implies
that
\begin{align}
	\mathbf{d}^{AP}_i = \mathbf{0}, 
	\quad {\rm for\;} i\in \lbrace 2,3,4 \rbrace\quad .
	\label{eq:apfact}
\end{align}

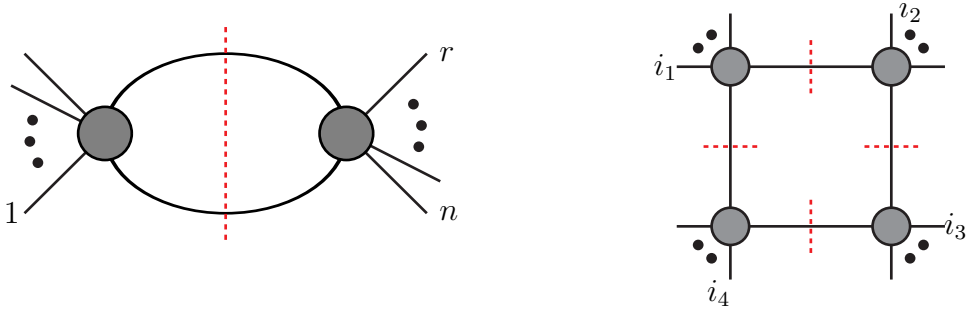
\begin{figure}[ht]
\centerline{
    \begin{picture}(230,170)(-95,-95)    
	    \SetWidth{1}
	    \GOval(0,0)(30,45)(0){1}
	    \Line(45,0)(75,30)
	    \Line(45,0)(75,-30)
	    \Line(45,0)(80,-18)
	    \Vertex(-72,5){2}
	    \Vertex(-73,-3){2}
	    \Vertex(-70,-11){2}
	    \Vertex(72,-5){2}
	    \Vertex(73,3){2}
	    \Vertex(70,11){2}
	    \Line(-45,0)(-80,18)
	    \Line(-45,0)(-75,30)
	    \Line(-45,0)(-75,-30)
	    \GOval(45,0)(10,10)(0){0.5}
	    \GOval(-45,0)(10,10)(0){0.5}
	    \Text(-80,-30)[c]{$1$}
	    \Text(80,30)[l]{$r$}
	    \Text(80,-30)[l]{$n$}
	    \SetColor{Red}
	\DashLine(0,40)(0,-40){2}
    \end{picture}
    \begin{picture}(140,100)(-50,-60)    
	    \SetWidth{1}
     \Line( 0, 0)( 0,60)
     \Line( 0,60)(60,60)
     \Line(60,60)(60, 0)
     \Line(60, 0)( 0, 0)
     \Line( 0, 0)(-20,0)
     \Line( 0, 0)(0,-20)
     \Line(60,60)(60,80)
     \Line(60,60)(80,60)
     \Line(0,60)(0,80)
     \Line(0,60)(-20,60)
     \Line(60, 0)(80,0)
     \Line(60, 0)(60,-20)
   \CCirc(60,60){7}{Black}{Gray} 
   \CCirc(0,60){7}{Black}{Gray} 
   \CCirc(60,0){7}{Black}{Gray} 
   \CCirc(0,0){7}{Black}{Gray} 
     \Text(0,-25)[r]{${i_4}$}   
     \Text(-20,60)[r]{${i_1}$}   
     \Text(63,82)[l]{${i_2}$}  
     \Text(80,0)[l]{${i_3}$}   
	    \Vertex(-12,-7){2}
	    \Vertex(-7,-12){2}
	    \Vertex(72,-7){2}
	    \Vertex(67,-12){2}
	    \Vertex(-12,67){2}
	    \Vertex(-7,72){2}
	    \Vertex(72,67){2}
	    \Vertex(67,72){2}
    \SetColor{Red}
     \DashLine(30,-10)(30,10){2}
     \DashLine(30,50)(30,70){2}
     \DashLine(-10,30)(10,30){2}     
     \DashLine(50,30)(70,30){2}
    \end{picture} 
   } 
	\caption[Cuts]{Old-fashioned and generalised unitarity: 
	the two-particle cut on the left are a consistency
	condition on the final amplitude, which can be used to
	constrain its form;
	the generalised four-particle cut on 
	the right allows the direct 
	computation of the box 
	coefficient $d^{[i_1,i_2,i_3,i_4]}_4$.
	}
    \label{fig:4dcuts}
\end{figure}

We adopt the Grassmann delta-function
$\delta^{(8)}\left(|i\ra \eta_{iA}\right)$ 
notation~\cite{Bianchi:2008pu}, which bundles together the 
states in the supermultiplet as related by
supersymmetric (SUSY) Ward identities.
Negative helicity gluon states are extracted by applying the
usual functional derivatives ${\delta^4 \over \delta^4\eta_i}$; 
in practice this amounts simply to partial derivatives and
then setting all $\eta$ to 0.
This allows us to deal with the cyclically invariant MHV 
superamplitude, and project out gluon amplitude on the RHS 
of equation~\bref{eq:conj} with
\begin{align}
	A_n^{\mathcal{N}=4}\left( 1^+,...,i^-,..,j^-,...,n^+
	\right)={\delta^4 \over \delta\eta^4_i}	
	{\delta^4 \over \delta\eta^4_j}	
	\left[ A_n^{\rm MHV}\right]\quad .
\end{align}
In this broader framework,
two-particle unitarity cuts depicted on the left 
of figure~\ref{fig:4dcuts}, 
can be expressed as a product of two
MHV superamplitudes. 
Considering, for example, the $q^2_{1r}$-channel cut, 
\begin{align}
	A^{\rm MHV}_n\biggr|_{q_{1r} \;{\rm cut}}=&
	\int d^4\eta_{\ell_1}d^4\eta_{\ell_r} {
		\delta^{(8)}\left( L\right)\over \la
	\ell_1 1\ra\la 12\ra \cdots \la (r-1)\ell_r\ra 
	\la \ell_r\ell_1\ra }\times
	{\delta^{(8)}\left( R\right)\over 
	\la \ell_r r\ra\cdots \la n\ell_1\ra \la \ell_1\ell_r\ra } ,
	\notag \\
	&L\equiv |i\ra 
		\eta_{iA} ,\;i\in \lbrace \mi \ell_1,1,
		...,r-1,\ell_r\rbrace\; ;\; \\
	&R\equiv |i\ra 
		\eta_{iA} ,\;i\in \lbrace \mi \ell_1,1,
		...,r-1,\ell_r\rbrace \; . \nonumber
	\label{eq:mhvcut}
\end{align}
The left-hand delta function can be made independent of 
the integral variables on the support of the right-hand
one, by
carrying out the Grassman integration
\begin{align}
	A^{\rm MHV}_n\biggr|_{q_{1r} \;{\rm cut}}=&
	 \quad {\delta^{(8)}\left(L+R\right)\over \la
	\ell_1 1\ra\la 12\ra \cdots \la (r-1)\ell_r\ra 
	}
	\times{\la \ell_1\ell_r\ra^2\over 
	\la \ell_r r\ra\cdots \la n\ell_1\ra  }& 
	\notag \\
	= 
	 &{\delta^{(8)}\left(|i\ra 
		\eta_{iA}\right) \over 2\la 12...n1\ra}\; 
		\left[{\tr(1q_{2(r-1)}(r-1)q_{r1})\over
			\ell_{2}^2\ell_{r\mi 1}^2} 
	+{\tr(nq_{1r}(r)q_{(r+1)n})\over
			\ell_{n}^2\ell_{r+ 1}^2}
			\right] \; ,
	\notag \\
	& i\in \lbrace 1,2,...,n
	 \rbrace \; .
\end{align}
We are then left only with box 
contributions from four-dimensional 
cuts.

We can more conveniently use generalised 
unitarity~\cite{Britto:2004nc} to compute $\mathbf{d}_4$.
Each box coefficient can be computed by imposing the four dimensional on-shell conditions
\begin{align}
	\ell^2_{i_k}= (\ell_1-q_{1i_k})^2 = 0,\; i\in {1,2,3,4}
	\label{eq:4dcc}
\end{align}
resulting in  the product of amplitudes on the generalised cut 
depicted on  the right of figure~\ref{fig:4dcuts}. 
The coefficient should be averaged over the two 
solutions of the constraints in equation~\bref{eq:4dcc}.
\begin{align}
	d_4^{[i_1,i_2,i_3,i_4]} = {1\over 2}\sum_{l_\pm}
	&A^{\rm tree}\left(\mi \ell_{i_1},i_1,...
					,i_2-1,\ell_{i_2}\right)
	\times
	A^{\rm tree}\left(\mi \ell_{i_2},i_2,...
					,i_3-1,\ell_{i_3}\right)
	\notag \\
	\times &
	A^{\rm tree}\left(\mi \ell_{i_3},i_3,...
					,i_4-1,\ell_{i_4}\right)
	\times
	A^{\rm tree}\left(\mi \ell_{i_4},i_4,...
					,i_1-1,\ell_{i_1}\right)
	\quad .
	\label{eq:4d4a}
\end{align}

The only cut configuration with non-vanishing amplitudes composing
it involves opposite facing 3-point amplitudes,
which upon evaluation gives
\begin{align}
	d_4^{[i_1,i_1+1,i_3,i_3+1]}= {\delta^{(8)}
	(|i\ra \eta_{iA})\over
	\la 12...n1\ra} \times \tr(i_1q_{i_1+1,i_3}i_3q_{i_3+1,i_1})\,.
	\label{eq:a4mhv}
\end{align}
Thus the amplitude is
\begin{align}
	A_n^{\rm MHV} = {1\over 4}{\delta^{(8)}
	(|i\ra \eta_{iA})
	\over \la 12...n1\ra}
	\sum_{i_1,i_3=1}^n
	\tr(i_1q_{i_1+1,i_3}i_3q_{i_3+1,i_1})
	I_4^{[i_1,i_1+1,i_3,i_1+1]}
	+\mathcal{O}(\epsilon)\; ,
	\label{eq:n4trunc}
\end{align}
where $I_4$ is in $4-2\epsilon$ dimensions.
The terms $\mathcal{O}(\epsilon)$ are 
(implicitly) determined in sections~\ref{sec:cuts} 
and~\ref{sec:gencut}.

The AP amplitude manifests a structure which is somewhat converse; 
its four-dimensional unitarity cuts vanish:
any on-shell QCD amplitude inserted into the 
cut diagrams of figure~\ref{fig:4dcuts}
with
all-plus helicity external legs vanishes.
It is, however, highly constrained and as such 
the all-$n$ form was determined ad hoc by 
 symmetry principles and limiting 
 behaviour~\cite{Mahlon:1993si,Bern:1993qk}.
 \begin{align}
	 A_n^{\rm AP}= \sum_{1\leq i_1<i_2<i_3<i_4\leq n}
	 {\tr_-(i_1i_2i_3i_4)\over \la 12...n1 \ra}
	 +\mathcal{O}(\epsilon)\quad .
	 \label{eq:apfin}
 \end{align}
Equation~\bref{eq:apfin} can be recovered 
from equations~\bref{eq:a4},
\bref{eq:a5}, and \bref{eq:a6} by noting that
\begin{align}
	\epsilon(1-\epsilon)I_4 &= {1\over 6}+\mathcal{O}(\epsilon)
	\; ,
	\notag \\
	\epsilon(1-\epsilon)I_5 &= {1\over 24}+\mathcal{O}(\epsilon)
	\quad .
\end{align}

The two sides of the conjecture~\bref{eq:conj} were computed in
different ways by BDDK:
the all-epsilon structure of MHV amplitude 
was computed using the string-based formalism
whereas AP amplitude, was computed with 
a prototypical version of $D$-dimensional unitarity.
This is effectively equivalent to taking unitarity cuts with
massive on-shell states, specifically applying 
the techniques of Bern
and Morgan~\cite{Bern:1995db}
 to compute the expressions in
equations~\bref{eq:a4}, \bref{eq:a5} and \bref{eq:a6};
a technique made especially simple
thanks to the equivalence in equation~\bref{eq:apscal}. 

Moreover, as BDDK highlight, 
the $D$-dimensional unitarity technique provides  
the full structure of the amplitude and can thus 
be made use of to prove the conjecture. 
An alternative statement to~\bref{eq:conj} in terms of cuts is thus
\begin{align}
	A^{\rm AP}_n\biggr|^{\mu^2\neq 0}_{q_{rs}\; {\rm cut}} = A^{\mathcal{N}=4}
	(1^+,2^+,\ldots,i^-,\ldots,j^-,\ldots,n^+)
	\left[{2\mu^4\over \la ij\ra^4}\right]\biggr|^{\mu^2\neq 0}_{q_{rs}\; {\rm cut}}
	\; ;
	\label{eq:cutproof1}
\end{align}
for any $q_{rs}^2$ channel.
Here $\mu^2$ is a mass-like parameter associated with the 
$(\mi2\epsilon)$-dimensional component of the loop integral.
The statement~\bref{eq:cutproof1} is proved in section~\ref{sec:cuts} thanks to massive spinor-helicity
formalisms~\cite{Dittmaier:1998nn,Arkani-Hamed:2017jhn} and 
insights into supersymmetric Ward identities~\cite{Boels:2011zz}
for Coulomb-branch 
amplitudes~\cite{Craig:2011ws,Kiermaier:2011cr,Elvang:2011ub}
in particular an axial-gauge choice, which is equivalent to a 
choice of basis with which to define the intermediate states in
the massive unitarity cut.

\begin{figure}[ht]
\centerline{
\begin{picture}(200,170)(-100,-100)    
	    \SetWidth{2}
	    \GOval(0,0)(30,45)(0){1}
	    \SetWidth{1}
	    \Line(45,0)(75,30)
	    \Line(45,0)(75,-30)
	    \Line(45,0)(80,-18)
	    \Vertex(-72,5){2}
	    \Vertex(-73,-3){2}
	    \Vertex(-70,-11){2}
	    \Vertex(72,-5){2}
	    \Vertex(73,3){2}
	    \Vertex(70,11){2}
	    \Line(-45,0)(-80,18)
	    \Line(-45,0)(-75,30)
	    \Line(-45,0)(-75,-30)
	    \GOval(45,0)(10,10)(0){0.5}
	    \GOval(-45,0)(10,10)(0){0.5}
	    \Text(-80,-30)[c]{$1$}
	    \Text(80,30)[l]{$r$}
	    \Text(80,-30)[l]{$n$}
	    \SetColor{Blue}
	\DashLine(0,40)(0,-40){5}
    \end{picture}
    }

\centerline{
\begin{picture}(140,100)(0,-34)    
	    \SetWidth{2}
     \Line( 0, 0)( 0,60)
     \Line( 0,60)(60,60)
     \Line(60,60)(60, 0)
     \Line(60, 0)( 0, 0)
	    \SetWidth{1}
     \Line( 0, 0)(-20,-20)
     \Line(60,60)(80,80)
     \Line(0,60)(0,80)
     \Line(0,60)(-20,60)
     \Line(60, 0)(80,0)
     \Line(60, 0)(60,-20)
   \CCirc(60,60){7}{Black}{Gray} 
   \CCirc(0,60){7}{Black}{Gray} 
   \CCirc(60,0){7}{Black}{Gray} 
   \CCirc(0,0){7}{Black}{Gray} 
     \Text(-8,-25)[r]{${i_4}$}   
     \Text(-20,60)[r]{${i_1}$}   
     \Text(70,87)[l]{${i_2}$}  
     \Text(80,0)[l]{${i_3}$}   
	    \Vertex(72,-7){2}
	    \Vertex(67,-12){2}
	    \Vertex(-12,67){2}
	    \Vertex(-7,72){2}
    \SetColor{Blue}
     \DashLine(30,-10)(30,10){5}
     \DashLine(30,50)(30,70){5}
     \DashLine(-10,30)(10,30){5}     
     \DashLine(50,30)(70,30){5}
\end{picture} 
\begin{picture}(100,110)(-50,-70)    
	\Line(-30,-40)(-55,-46)
		\Vertex(-44,-49){2}
		\Vertex(-38,-55){2}
	\Line(-30,-40)(-35,-65)
	\Text(-28,-60)[l]{$i_1$}   
	\Line(30,-40)(55,-46)
		\Vertex(44,-49){2}
		\Vertex(38,-55){2}
	\Line(30,-40)(35,-65)
	\Text(59,-40)[r]{$i_5$}   
	\Line(-65,18)(-45,5)
	\Line(-65,-8)(-45,5)
		\Vertex(-60,0){2}
		\Vertex(-60,10){2}
	\Text(-64,-10)[r]{$i_2$}   
	\Line(65,18)(45,5)
	\Line(65,-8)(45,5)
		\Vertex(60,0){2}
		\Vertex(60,10){2}
	\Text(69,18)[l]{$i_3$}   
	\Line(14,60)(0,40)
	\Line(-14,60)(0,40)
	\Vertex(-5,57){2}
	\Vertex(5,57){2}
	\Text(-16,58)[r]{$i_4$}   
\SetWidth{2}
	\Line(-30,-40)(30,-40)
	\Line(45,5)(30,-40)
	\Line(-30,-40)(-45,5)
	\Line(0,40)(-45,5)
	\Line(45,5)(0,40)
\SetWidth{1}
   \CCirc(-30,-40){7}{Black}{Gray} 
   \CCirc(30,-40){7}{Black}{Gray} 
   \CCirc(45,5){7}{Black}{Gray} 
   \CCirc(-45,5){7}{Black}{Gray} 
   \CCirc(0,40){7}{Black}{Gray} 
	    \SetColor{Blue}
		\DashLine(0,-50)(0,-30){5}
		\DashLine(-50,-20)(-25,-11){5}
		\DashLine(50,-20)(25,-11){5}
		\DashLine(-30,34)(-11,14){5}
		\DashLine(30,34)(11,14){5}
    \end{picture} 
   } 
	\caption[DCuts]{
		TOP: The $D$-dimensional cuts capture the full
		amplitude, it is in this context that the 
		BDDK conjecture is proved in section~\ref{sec:cuts}.

		BOTTOM: Generalised $D$-dimensional cuts can be used
		to compute the functional form of the all-plus,
		and thus the MHV through the relation~\bref{eq:conj}.
	}
    \label{fig:Ddcuts}
\end{figure}
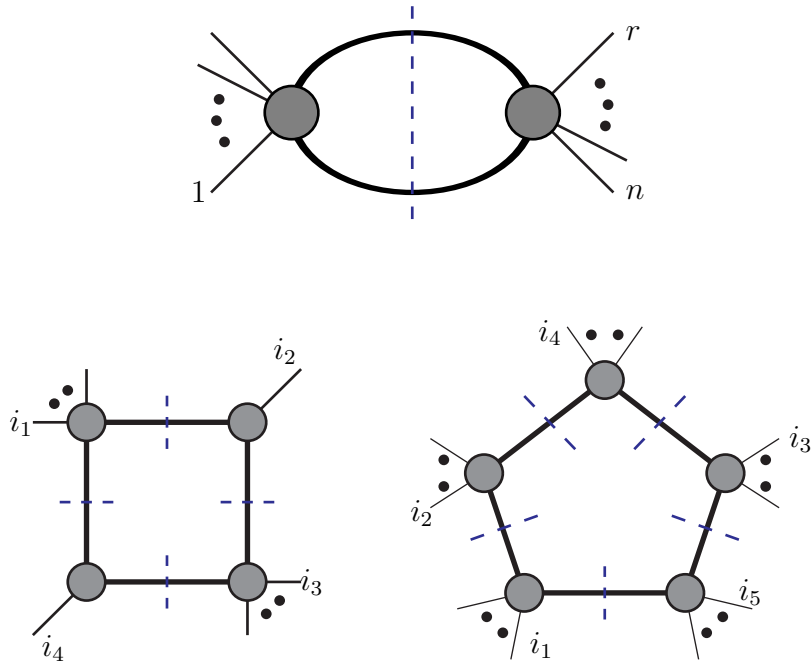

In section~\ref{sec:gencut} an ansatz is presented for 
the all-$n$ all-$\epsilon$ form
for the AP (and thus through the dimension shift in 
equation~\bref{eq:conj}	the $\mathcal{N}=4$ 
MHV) amplitude through the computation of 
generalised $D$-dimensional cuts, depicted in figure~\ref{fig:Ddcuts}.
This method is directly analogous to 
the one first introduced for four-dimensional cuts 
in~\cite{Britto:2004nc};
it produces a compact result thanks to the simple form of a solution to the
on-shell conditions for a given pentagon,
$I_5^{[i_1,i_2,i_3,i_4,i_5]}
$\begin{align}
	\ell_{i_1}^\nu&= -{\tr_5\left(q_{i_1i_2}q_{i_2i_3}
		q_{i_3i_4}q_{i_4i_5}\gamma^\nu\right)\over
		2\tr_5\left(q_{i_1i_2}q_{i_2i_3}
		q_{i_3i_4}\right) } \quad .
\end{align}
We then conclude with some observations and outlook.

\newpage

%% file: proof-sec.tex

\section{Proof of the conjecture}
\label{sec:cuts}
As stated in the original discussion in~\cite{Bern:1996ja},
an equivalent statement to the conjecture~\bref{eq:conj}
is that the $D$-dimensional cuts
match:
\begin{align}
	A^{AP}\biggr|^{\mu^2\neq 0}_{q_{rs}\;{\rm cut}} 
	= A^{\mathcal{N}=4}
	(++ \cdots i^- \cdots j^- \cdots ++)
	\left[{2\mu^4\over \la ij\ra^4}\right]
	\biggr|^{\mu^2\neq 0}_{q_{rs}\;{\rm cut}}
	\quad  .
	\label{eq:cutproof2}
\end{align}
It is in showing this relationship that we prove this conjecture
here.

\subsection{Unitarity cuts}
It has long been known that unitarity cuts provide tight constraints on 
the functional form of amplitudes~\cite{Eden1966} and
can be used as a tool to compute them~\cite{Bern:1994zx,Bern:1994cg}.
A technique emblematic of contemporary on-shell methods for
computing scattering amplitudes,
generalised unitarity~\cite{Bern:1994zx,Bern:1994cg,Britto:2004nc,
Britto:2006sj,Forde:2007mi,Dunbar:2009ax,Mastrolia:2009dr}, 
is able to fully fix box, triangle and bubble coefficients 
in supersymmetric theories at leading order in epsilon.
However, rational terms, in particular 
those that compose the all-plus and single-minus
amplitudes, are non-vanishing in non-supersymmetric theories;
supplementary approaches~\cite{Bern:2005hs,Badger:2008cm,
Dunbar:2010wu,Alston:2012xd,Dunbar:2017nfy}
are needed to overcome these limitations.

A more complete unitarity technique  
known as ``$D$-dimensional" unitarity~\cite{Giele:2008ve,Ellis:2008ir,Ossola:2008xq,Badger:2008cm} 
extends the conventional version, but considers additional singularities of the 
integrand which capture the full singular 
structure of the amplitude;
these extra singularities can be counted when the ``on-shell"
states of the unitarity cut are considered to be
$D=4-2\epsilon$ dimensional.

After formally splitting the $D$-dimensional loop momentum $\ell$ into a 
four-dimensional component and an orthogonal ($-2\epsilon$)-dimensional one,
\begin{align}
	\ell = l^\mu + \ell^{[\mi 2\epsilon]}
\end{align}
the cut condition becomes
\begin{align}
	\ell^2 = l^2 - \mu^2 = 0 \; ,
\end{align}
where we have defined 
\begin{align}
\mu^2 \equiv \left(\ell^{[\mi 2\epsilon]}\right)^2 \,.
\end{align}
The integral measure can then be transformed according to
\begin{align}
	d^{4\mi 2\epsilon}\ell 
	&= -{d\mu^2 \over (-\mu^2)^{1+\epsilon}}d^4l
	\notag \\
	&= {d\mu^2 \over \mu^2}d^4l+\mathcal{O}(\epsilon)
	\label{eq:seppo} \quad .
\end{align}
The $\mu^2=0$ pole captures the four-dimensional structure,
whereas the singular locus for the combination of all cut conditions
\begin{align}
	\ell_i^2 = l_i^2 - \mu^2 = 0 
\end{align}
encodes all-epsilon properties with coefficients characterised 
by the ``off-shell structure" of the tree-level theory.
Thus $\mu^2$ can be considered as an extra parameter subject to 
the cut constraints, and is ultimately the mass term in the 
tree-level amplitudes which are used to fix the 
integral coefficients.

As with four-dimensional unitarity, a
set of $k$ massive-cut conditions 
singles out a codimension-$k$ singular locus in the 
domain of integration;
the extra integral parameter $\mu^2$ puts the dimension of this 
domain, and thus the number of constraints on the maximal cut,
at $k=5$ as opposed to $k=4$ in the massless (or $\mu^2=0$) case.
Thus for a $k$-line cut, $k-5$ free
parameters are unfixed by the constraints. 
The coefficients can be determined from these cut expressions
through
taking careful limits depending on the 
residues coming from poles 
in the unfixed parameters~\cite{Forde:2007mi,Badger:2008cm}.
How this is done is contingent on a choice of integral basis:
both the AP and MHV amplitudes
 are free of bubble and triangle contributions,
so just
the $\mu^2$ degree of freedom need be considered to capture box 
coefficients.
The discussion of what basis to use 
in computation of the amplitudes is carried out 
in section~\ref{sec:gencut}.

Four-dimensional cuts constrain an amplitude by 
requiring that any two-particle cuts reproduce a product 
of tree-level amplitudes 
\begin{align}
	A_n\biggr|^{\mu^2= 0}_{q_{rs}\;{\rm cut}}=&
	\sum_{h_s,h_r } 
	A^{\tree}_{r+2}(\mi l_r^{h_r}, 
	r,r+1,...,s- 1,l_{s}^{\mi h_{s}},)
\times	
	A^{\rm tree}_{n-r+2}(\mi l_{s}^{\mi h_s},
	,s,s+1,...r-1,l_{r}^{ h_{r}})
	\quad .
	\label{eq:focut}
\end{align}
This can be fit into a broader framework
by noting that the eigenstates of the little group for massless
particles, $U(1)$, are represented by the sum over helicities.
The helicity ``weight" which induces a helicity weighting
(or complex phase)
for each particle cancels across the cut; this is
how the cut can be viewed as a $U(1)$ contraction.
In theories with rational terms, cuts of the form~\ref{eq:focut}
are incomplete, as a consequence of the intermediate states
living in $D=4-2\epsilon$ dimensions.
Taking a $D$-dimensional unitarity requires
a concrete scheme for defining these states crossing the cut.
This has previously been done using 
six-dimensional spinor 
helicity~\cite{Cheung2009,Bern2010,Badger2017}, but
here we propose simply using off-shell/equal-mass amplitudes.

Equation~\ref{eq:seppo} provides a natural 
motivation to use massive amplitudes\footnote{The fact that 
Pauli-Villars regularisation~\cite{Pauli:1949zm} functions
by introducing masses also 
supports this.}.
An heuristic justification for this is the fact renormalisation and
resolution of infrared singularities always introduces an external
scale, and thus off-shell expressions of the classical theory
will be needed to fully define a loop amplitude.
The massive spinor-helicity 
formalism of Arkani-Hamed, Huang and 
Huang~\cite{Arkani-Hamed:2017jhn} (AHH) provides
a contemporary framework for combining these expressions.

In the AHH formalism, a two-particle unitarity cut 
consists of a sum over several inner products each 
carried out over the particles crossing the cut 
\begin{align}
	A_n\biggr|^{\mu^2\neq 0}_{q_{rs}\;{\rm cut}}=
	\sum_{\rm particles}\left\langle 
	A^{\rm tree}_{L}(\mi{\bm l}_r^{\sigma_r}, r,r+1,...,s-1 ,
	{\bm l}_{s}^{\sigma_{s}})
	,
	A^{\rm tree}_{R}(\mi
	{\bm l}_{r}^{\sigma_{s}},s,...,r-1,{\bm l}_r^{\sigma_r})
	\right\rangle
	\; ,
	\label{eq:inprod}
\end{align}
where the boldface notation indicates suppressed, 
symmetrised $SU(2)$ indices, 
the little group of massive particles in 3+1 dimensions.
The inner product contracts these indices, and 
the contraction has a clear interpretation in terms of 
the eigenstates of the $SU(2)$ little group of the particles
of spin $\sigma$ which ``cross" the cut.
The AHH formalism thus represents
this product as rank--$(2\sigma_s+2\sigma_r)$ tensors of $SU(2)$
contracted with two-dimensional 
Levi-Civita symbols $\epsilon_{\alpha^r_j\beta^r_j}$ and $\epsilon_{\alpha^s_k\beta^s_k}$,
\begin{align}
	A_n\biggr|^{\mu^2\neq 0}_{q_{rs}\;{\rm cut}}=
	\sum_{\rm particles} 
	\quad  A_{L}^{\alpha^r_{1}...\alpha^r_{2\sigma_r};\alpha^s_{1}...\alpha^s_{2\sigma_s}}
	\prod_{j=1}^{2\sigma_r}\left[\epsilon_{\alpha^r_j\beta^r_j}
	\right]
	\prod_{k=1}^{2\sigma_s}\left[\epsilon_{\alpha^s_k\beta^s_k}
	\right]
	A_{R}^{\beta^s_1...\beta^s_{2\sigma_s};\beta^r_1...\beta^r_{2\sigma_r}}
	\,.
	\label{eq:masscut}
\end{align}

There is also a constraint on the expressions which feed into the
cut expression~\bref{eq:inprod}: the quantum states in a given
channel must be consistently defined, i.e. defined in the same
frame of reference.
In practice this is simply done by fixing an axial gauge 
vector\footnote{This is most often denoted as $q$ in the literature.
It is equivalent to fixing a Lorentz frame and thus is also 
referred to as a ``$q$-frame" in eg.~\cite{Kiermaier:2011cr}.
Note also that choosing $\gg$ to be complex is also an option.}, 
which we term $p_\chi$, against which to define helicities, or 
spin states of massive particles.

To use these tools to prove the formula~\bref{eq:cutproof2} we 
require the tree-level expressions.

\subsection{Tree-level off-shell amplitudes}
\label{sec:trees}
The all-multiplicity tree-level
expressions that need to be fed into equation~\bref{eq:inprod}
have been known for a long time and exist
in multiple forms throughout the 
literature~\cite{Forde:2005ue,Rodrigo:2005eu,Ferrario:2006np,Kiermaier:2011cr,Ochirov2018},
mostly in the older massive spinor-helicity formalism
of Dittmaier~\cite{Dittmaier:1998nn}.
As with the tree amplitudes of all-massless particles, 
compact expressions for massive amplitudes are neatly 
related to each other through SUSY 
Ward identities~\cite{Schwinn:2006ca,Boels:2011zz}.
These can be bundled together in a super-amplitude 
like the massless-MHV case, with a clear interpretation 
in terms of taking $\mathcal{N}=4$ away from the origin
of the moduli space~\cite{Craig:2011ws,Kiermaier:2011cr},
where
the origin corresponds to the massless 
theory\footnote{The Coulomb branch can also be obtained
by compactifying the massless $\mathcal{N}=(1,1)$ 
maximal theory in 6 dimensions~\cite{Cheung2009,Bern2010}.}. 
The result is a super-amplitude in what is 
termed ``MHV band"~\cite{Craig:2011ws}.
The MHV-band superamplitude differs from
the massless MHV super-amplitude in that 
the delta functions are not homogeneous in the $\eta$
variables which carry the helicity weight.
We discuss how to interpolate between the two formalisms
and collect the necessary expressions here. 

The AHH massive spinor-helicity formalism~\cite{Arkani-Hamed:2017jhn} 
deployed here extends 
the compactness of conventional spinor helicity by  
defining a given amplitude with massive external particles
$1$ and $n$, of spin $\sigma_1$ and $\sigma_n$ respectively, as a rank 
$2\sigma_1+2\sigma_n$ tensor of $SU(2)$, the little group for 
massive particles:
\begin{align}
	A^{\rm tree}_n(\mathbf{1}^{\sigma_1},2^{h_2},...,
				(n-1)^{h_{n\mi 1}},
			\mathbf{n}^{\sigma_n} )\quad .
\end{align}
In practice, to compute a cut, we will need to choose a basis
and express the tree-level amplitudes on the right hand side of
equation~\bref{eq:masscut} in the form
\begin{align}
	A^{\rm tree}_n(\mathbf{1}^{\sigma_1},2^{h_2},...,
	\bar{\mathbf{n}}^{\sigma_n} )
	= \sum_{h_1,h_n} 
	A_n^{h_{1}h_{n}}
	{\bm\zeta}^{h_1}{\bm\zeta}^{h_2}\; ;
	\label{eq:zbas}
\end{align}
here, on the LHS $\sigma_i$ ($h_i$) is the spin (helicity) of the
massive (massless) particle; on the RHS we use indices $h_1, h_n$
to label the spin states of the massive particles in the chosen 
basis.
{
The boldface variables 
${\bm\zeta}\equiv \prod_{k=1}^{2\sigma}\zeta^{\alpha_k}$ 
again carry suppressed 
$SU(2)$ tensor indices $\alpha_k$.
Each tensor indexes the spin-states of 
the massive particles in a given frame of reference,
or more specifically relative to a choice of axial gauge vector,
which we term $p_\chi$. 
These spinning states are pseudo-orthonormal,
\begin{align}
	\la {\bm \zeta}^{ h_a} , 
	{\bm \zeta}^{- h_b} \ra 
	= \left( \pm 1\right)^{2h_a} \delta_{ab}\, ,
\end{align}
meaning the product in equation~\bref{eq:inprod} can be
expressed as 
a sum over products of complex valued functions in 
a given $p_\chi$ basis.
As demonstrated by AHH in~\cite{Arkani-Hamed:2017jhn},
the high-energy limits ($m^2\rightarrow 0$)
of the coefficients $A_n^{h_{1}h_{n}}$ are
the tree amplitudes of massless particles.
}

AP amplitudes must satisfy the constraint
in equation~\bref{eq:masscut};
thanks to all-$n$ expressions for all-plus helicity
gluons with two other massive particles,
the AP amplitude can
be reconstructed by the product of scalar amplitudes that was
exploited
in~\cite{Bern:1996ja}
\begin{align}
	A^{\rm AP}_n
	\biggr|^{\mu^2\neq 0}_{q_{rs}\; {\rm cut}}=& 
	2\left\langle A^{\rm tree}_{L}(\mi {\bm l}_r^{0}, 
	r,r+1,...,s-1,
	{\bm l}_{s}^{0})
	,
	A^{\rm tree}_{R}(\mi{\bm l}_{s}^{0},
	,s,s+1,...,r-1,{\bm l}_{r}^{0})\right\rangle
	\notag \\
	=& 
	2A^{\rm tree}_{L}(\mi {\bm l}_r^{0}, 
	r,r+1,...,s-1,
	{\bm l}_{s}^{0})\times	
	A^{\rm tree}_{R}(\mi{\bm l}_{s}^{0},
	,s,s+1,...,r-1,{\bm l}_{r}^{0})
	\; ,
	\label{eq:scalca}
\end{align}
as is consistent with the string-derived equation~\bref{eq:apscal}.
In the scalar case, the amplitudes are scalar
functions thus reducing the inner product to conventional 
multiplication. 

The original computations 
which resulted in equations
\bref{eq:a4}-\bref{eq:a6} reconstructed integrand 
polynomials in the parameter $\mu^2$ and entailed 
finding a function which was consistent with it. 
In section~\ref{sec:gencut} we will adopt a different and 
more direct approach, which builds on the work of 
Badger~\cite{Badger:2008cm} to directly extract the coefficients
of a scalar integral in a simple manner.
It is the functions $A_n^{\rm tree}$ needed as 
input for the cut constraints in equation~\bref{eq:masscut} which 
are the focus of this section.
To evaluate the cut AP amplitudes
we use the all-multiplicity expressions for tree-level amplitudes
of positive-helicity gluons and two equally massive scalars,
first computed by Forde and Kosower 
through recursion relations~\cite{Forde:2005ue}:
\begin{align}
	A^{\rm tree}&(\mathbf{1}^0,
	2^+,3^+,...,(n-1)^+,\mathbf{n}^0)=
	& {-
		\sum_{j=1}^{\left\lfloor {n\over 2}\right\rfloor - 1}\kappa_j 
		(-m^2)^j\over (s_{12}-m^2)\la 23 ...(n\mi 2)(n\mi 1)\ra (s_{n-1,n}-m^2)}
				\label{eq:ap2s}
\end{align}
with
\begin{align}
	\kappa_j = \sum_{\lbrace w_i \rbrace_{i=1}^{j-1}}^{n-3}
	{[2|1q_{2w_1}w_1q_{w_1w_2}w_2\cdots w_{j-1}q_{w_{j-1}w_j}|n-1]
	\over \prod_{r=1}^{j-1}(q_{1(w_r)}^2-m^2)(q_{1w_r+1}^2-m^2)}
\end{align}
where the sets $\lbrace w_i \rbrace$ are defined by
the conditions\footnote{See also~\cite{Forde:2005ue}
for more explicit representations 
of the sum over $\lbrace w_j \rbrace$.}:
\begin{align}
       w_j = n-1,\quad w_0 = 1, \quad
	w_i=w_{i - 1}+2\quad .
\end{align}
	Other more compact representations of
$A^{\rm tree}_n(\mathbf{1}^0,2^+,3^+,...,(n-1)^+,\mathbf{n}^0)$
have also been 
computed~\cite{Rodrigo:2005eu,Ferrario:2006np,Kiermaier:2011cr,Ochirov2018}.
However, we would like a representation that is 
 arranged as a series in $m^2$ with only at worst
linear dependence on the massive momenta in the numerators, which helps to
insert cut solutions into amplitudes.
By means of the expression~\bref{eq:ap2s} for $n=4,5,6$,
BDDK were able to confirm the 
representations~\bref{eq:a4},~\bref{eq:a5}, and~\bref{eq:a6}
of the AP amplitudes.

The $\mathcal{N}=4$ super-multiplet is more complicated
to treat in this manner. We
 can begin by exploring the nature of off-shell amplitudes:
how does the cut~\bref{eq:masscut} look if quarks or vector bosons
were circulating in the loop?
As mentioned above, projection onto a state basis of the 
kind expressed in equation~\bref{eq:zbas} 
can be done by choosing a reference 
four-vector,
$\gg$, 
relative to which we define the spinning particles.
We build our basis using this reference, and interpolate between
old~\cite{Dittmaier:1998nn} and new~\cite{Arkani-Hamed:2017jhn} 
massive spinor-helicity formalisms.
This is ultimately equivalent to making a
choice of axial gauge~\cite{Schwinn2005}
and thus this choice is referred to simply as a gauge from here on.

Considering 
massive vector bosons $\mathbf{1}^1$ and $\mathbf{n}^1$ 
having opposite helicity,
it is straightforward to express the equivalent of
equation~\bref{eq:ap2s}: 
SUSY Ward 
identities~\cite{Schwinn:2006ca,Boels:2011zz} can 
be used to show that the
vector boson just 
involves the introduction of an overall factor
relative to the scalar case. 
The spin-one state coefficient of ${\bm \zeta}_1^-{\bm\zeta}_n^+$
in the expansion of equation~\bref{eq:zbas} is 
\begin{align}
	A^{-+}_n(\mathbf{1}^1,2^+,3^+,...,(n-1)^+,
	\mathbf{n}^1)
	=&  {\la\chi \lambda_1 \ra^2\over \la \chi \lambda_n\ra^2}
			A_n^{\rm tree}(\mathbf{1}^0,2^+,3^+,
			...,(n-1)^+,\mathbf{n}^0)	
			\label{eq:ap2vmp}
\end{align}
where $\la \chi |$ is a reference spinor
corresponding to the gauge choice and
$|\lambda_i \ra$ 
represents\footnote{Also denoted $|\lambda_i 
\ra\equiv | i^\flat\ra$ in~\cite{Schwinn2005}
or $| i^\perp \ra$ in~\cite{Craig:2011ws,Kiermaier:2011cr,Elvang:2011ub}.} the holomorphic spinor 
of the nullified momentum $p_{\lambda_i}$ defined such that
\begin{align}
	p_{\lambda_i} = p_i - {m^2\over [\chi|i|\chi\ra}p_\chi
	\quad .
\end{align}

We can extract the full rank $2\sigma_1+2\sigma_n$ 
tensor amplitude
by loosening the $\chi$ constraint, 
and considering how component amplitudes 
$A_n^{h_1h_n}$ in a particular gauge fix
the full $SU(2)$ tensor structure of the amplitude
through symmetry principles.
By fixing 
$\la \chi | \rightarrow \la \chi_n |$ where $\chi_n$ is defined 
such that
 $\la \chi_n \lambda_n\ra =m$ we get
\begin{align}
	A^{-+}_n(\mathbf{1}^1,2^+,3^+,...,(n-1)^+,
	\mathbf{n}^1)
	=&  {\la\chi_n \lambda_1 \ra^2\over m^2}
	A_n^{\rm tree}(\mathbf{1}^0,2^+,3^+,
			...,(n-1)^+,\mathbf{n}^0)	
				\;\quad .
				\label{eq:exp}
\end{align}
The spinor $\chi_n$ is a specific choice which 
favours the taking of the high-energy limit of 
tree expressions in the 
AHH scheme\footnote{In~\cite{Arkani-Hamed:2017jhn} they are 
denoted $\eta$ however we reserve the symbol $\eta$ to denote 
the Grassmann variables in the superamplitude.}~\cite{Arkani-Hamed:2017jhn}.
In general the momentum of each massive particle can 
be decomposed
\begin{align}
	\left(p_i\right)_{\dot{a}b} &= \left(p_{\lambda_i} 
	- p_{\chi_i}\right)_{\dot{a}b}=\lambda_{\dot{a}}^I
	\bar{\lambda}_{Ib} =  
	\notag \\
	\Rightarrow \lambda_{\dot{a}}^I &
	= \lambda_{\dot{a}}
	\left(\zeta^-\right)^I 
	+\chi_{\dot{a}}\left(\zeta^+ \right)^I\; ,
	\notag \\
	\bar{\lambda}_{b}^I &
	= \bar{\lambda}_{b}
	\left(\zeta^+\right)^I 
	+\bar{\chi}_{b}\left(\zeta^- \right)^I\; ,
\end{align}
and the spinors fixed such that 
$\la \lambda \chi\ra = [\lambda \chi ] = m $.
Thus from fixing the gauge $\chi\rightarrow \chi_n$ it is easy  
to induce
the gauge-covariant rank 4 amplitude from the $SU(2)$ symmetry
by promoting the coefficient 
of ${\bm \zeta}_1^-{\bm \zeta}_n^+$, the gauge-dependent 
factor in equation~\bref{eq:exp} can thus be generalised:
\begin{align}
	A_n^{\rm tree}(\mathbf{1}^1,2^+,3^+,...,(n-1)^+,
	\mathbf{n}^1)
	=& {\la\mathbf{1} \mathbf{n}\ra^2\over m^2}
	A_n^{\rm tree}(\mathbf{1}^0,2^+,3^+,
			...,(n-1)^+,\mathbf{n}^0)	
				\;\quad .
				\label{eq:ap2v}
\end{align}
The expression~\bref{eq:ap2v} can also be projected in
the frame of the massive particles spinning in the ``negative"
direction relative to their motion.
\begin{align}
	A_n^{--}(\mathbf{1}^1,2^+,3^+,...,(n-1)^+,
	\mathbf{n}^1)
	=&  {\la \lambda_1\lambda_n\ra^2\over m^2}
	A_n^{\rm tree}(\mathbf{1}^0,2^+,3^+,
			...,(n-1)^+,\mathbf{n}^0)	
				\;\quad ,
\end{align}
and from this it can be seen that
\begin{align}
	\lim_{m^2\rightarrow 0}
	\left[A_n^{--}(\mathbf{1}^1,2^+,3^+,...,(n-1)^+,
	\mathbf{n}^1)\right]
	= A_n^{\rm tree}(1^-,2^+,3^+,...,(n-1)^+,n^-)\; ,
\end{align}
providing a strong consistency check.

The quark version of~\bref{eq:ap2v}
has also been computed~\cite{Ferrario:2006np,Ochirov2018} and is simply 
\begin{align}
	A^{\tree}_n(\mathbf{1}^{1\over 2},2^+,3^+,...,(n-1)^+,
	\mathbf{n}^{1\over 2})
	=&  {\la\mathbf{1} \mathbf{n}\ra\over m}
	A^{\rm tree}_n(\mathbf{1}^0,2^+,3^+,
			...,(n-1)^+,\mathbf{n}^0)	\quad .
			\label{eq:ap2q}
\end{align}

The simple relation between the
scalar, fermion and gluon form factors in equations \bref{eq:ap2s}, 
\bref{eq:ap2v} and \bref{eq:ap2q} can be understood simply as a 
consequence of supersymmetric 
Ward identities~\cite{Schwinn:2006ca,Boels:2011zz}:
they are in the highly constrained 
``ultra-helicity-violating" (UHV) sector~\cite{Craig:2011ws},
so called because the particles spin states
have one fewer negative helicity states\footnote{The configurations of equation~\bref{eq:ap2vmp} 
would correspond to a single-minus amplitude, and indeed vanishes in the $m^2\rightarrow 0$ limit.}
than the MHV helicity configuration.

The fact that the full tensor-amplitude~\bref{eq:ap2v} produces the MHV case for free
is a feature of the blurring of the normally clear demarcation 
between helicity states in
the massive/spin case, unlike
the simpler massless/helicity case. 
As with the massless case, however,
these amplitudes can all be bundled into a super-amplitude of
$\mathcal{N} =4$ taken away from the origin of the moduli space
on the Coulomb branch:
the ``MHV-band" amplitudes read thus~\cite{Craig:2011ws}
\begin{align}
	A_{\rm tree}^{\rm MHV-band} = 
	{[\lambda_n\lambda_1]^2	\delta^\chi_{12}\delta^\chi_{34}
				\over  m^2q_{n2}^4}
	A(\mathbf{1}^0,2^+,3^+,
			...,(n-1)^+,\mathbf{n}^0)\quad .
		\label{eq:mhvband}
\end{align}

The expression $\delta^\chi_{12}\delta^\chi_{34}$
is a nonhomogeneous polynomial in the helicity (or more precisely
spin-state) weight carrying
variables $\eta_{ia}$.
Component amplitudes corresponding to different external
states can be extracted from the 
$\delta^\chi_{12}\delta^\chi_{34}$
in an identical way to how MHV states can be extracted from
$\delta^{(8)}(|i\ra \eta_{ia})$ using functional derivatives.
As for the massive vector boson amplitude in equation~\bref{eq:ap2v},
gauge covariance implies that a 
single class of amplitude contains multiple spin states. 

The structure and construction techniques of Coulomb branch
super-amplitudes is covered extensively with the Dittmaier 
massive-spinor-helicity formalism in 
references~\cite{Craig:2011ws,Kiermaier:2011cr,Elvang:2011ub}, 
and we refer the reader to these for more background.
The explicit general $p_\chi$-gauge form of 
$\delta^\chi_{12}\delta^\chi_{34}$
 can be found in~\cite{Craig:2011ws},
but we give an explicit form after we fixing
$\gg \cdot q_{rs}=0$, which is key to the proof of
the conjecture in the following section.

\subsection{Proving the conjecture from the Coulomb branch}
\label{sec:proof}
To prove the conjecture we consider the individual contributions
of particles as expressed in equation~\bref{eq:pdec}.

\begin{figure}[h]
\centerline{
    \begin{picture}(230,110)(-95,-55)    
	    \SetWidth{1}
	    \GOval(0,0)(30,45)(0){1}
	    \Line(45,0)(75,30)
	    \Line(45,0)(75,-30)
	    \Line(45,0)(80,-18)
	    \Vertex(-72,5){2}
	    \Vertex(-73,-3){2}
	    \Vertex(-70,-11){2}
	    \Vertex(72,-5){2}
	    \Vertex(73,3){2}
	    \Vertex(70,11){2}
	    \Line(-45,0)(-80,18)
	    \Line(-45,0)(-75,30)
	    \Line(-45,0)(-75,-30)
	    \GOval(45,0)(10,10)(0){0.5}
	    \GOval(-45,0)(10,10)(0){0.5}
	    \Text(-80,-30)[r]{$+$}
	    \Text(-84,20)[r]{$+$}
	    \Text(-78,33)[r]{$+$}
	    \Text(77,33)[l]{$+$}
	    \Text(84,-20)[l]{$-$}
	    \Text(80,-33)[l]{$-$}
	    \Text(-4,38)[r]{$+$}
	    \Text(-4,-40)[r]{$+$}
	    \Text(4,38)[l]{$-$}
	    \Text(4,-40)[l]{$-$}
	    \Text(-60,-50)[r]{${\rm \bf a.}$}
	    \SetColor{Blue}
	    \DashLine(0,40)(0,-40){6}
    \end{picture}
    \begin{picture}(230,110)(-95,-55)    
	    \SetColor{Black}
	    \SetWidth{1}
	    \GOval(0,0)(30,45)(0){1}
	    \Line(45,0)(75,30)
	    \Line(45,0)(75,-30)
	    \Line(45,0)(80,-18)
	    \Vertex(-72,5){2}
	    \Vertex(-73,-3){2}
	    \Vertex(-70,-11){2}
	    \Vertex(72,-5){2}
	    \Vertex(73,3){2}
	    \Vertex(70,11){2}
	    \Line(-45,0)(-80,18)
	    \Line(-45,0)(-75,30)
	    \Line(-45,0)(-75,-30)
	    \GOval(45,0)(10,10)(0){0.5}
	    \GOval(-45,0)(10,10)(0){0.5}
	    \Text(-80,-30)[r]{$+$}
	    \Text(-84,20)[r]{$+$}
	    \Text(-78,33)[r]{$+$}
	    \Text(77,33)[l]{$+$}
	    \Text(84,-20)[l]{$-$}
	    \Text(80,-33)[l]{$-$}
	    \Text(-4,38)[r]{$+$}
	    \Text(-4,-40)[r]{$-$}
	    \Text(4,38)[l]{$-$}
	    \Text(4,-40)[l]{$+$}
	    \Text(-60,-50)[r]{${\rm \bf b.}$}
	    \SetColor{Blue}
	    \DashLine(0,40)(0,-40){6}
    \end{picture}
    }
	\centerline{
    \begin{picture}(230,110)(-95,-55)    
	    \SetColor{Black}
	    \SetWidth{1}
	    \GOval(0,0)(30,45)(0){1}
	    \Line(45,0)(75,30)
	    \Line(45,0)(75,-30)
	    \Line(45,0)(80,-18)
	    \Vertex(-72,5){2}
	    \Vertex(-73,-3){2}
	    \Vertex(-70,-11){2}
	    \Vertex(72,-5){2}
	    \Vertex(73,3){2}
	    \Vertex(70,11){2}
	    \Line(-45,0)(-80,18)
	    \Line(-45,0)(-75,30)
	    \Line(-45,0)(-75,-30)
	    \GOval(45,0)(10,10)(0){0.5}
	    \GOval(-45,0)(10,10)(0){0.5}
	    \Text(-80,-30)[r]{$+$}
	    \Text(-84,20)[r]{$+$}
	    \Text(-78,33)[r]{$+$}
	    \Text(77,33)[l]{$+$}
	    \Text(84,-20)[l]{$-$}
	    \Text(80,-33)[l]{$-$}
	    \Text(-4,38)[r]{$-$}
	    \Text(-4,-40)[r]{$-$}
	    \Text(4,38)[l]{$+$}
	    \Text(4,-40)[l]{$+$}
	    \Text(-60,-50)[r]{${\rm \bf c.}$}
	    \SetColor{Blue}
	    \DashLine(0,40)(0,-40){6}
    \end{picture}
    }
	\caption[mhv cuts]{Example helicity configurations 
	of \textbf{a}. \textbf{b}. and \textbf{c}.
	Only ${\rm \bf c.}$, the MHV$\times$MHV cut
	survives in the full $\mathcal{N}=4$ multiplet in
	$q_{rs}\cdot\gg$ gauge.}
    \label{fig:gcut}
\end{figure}
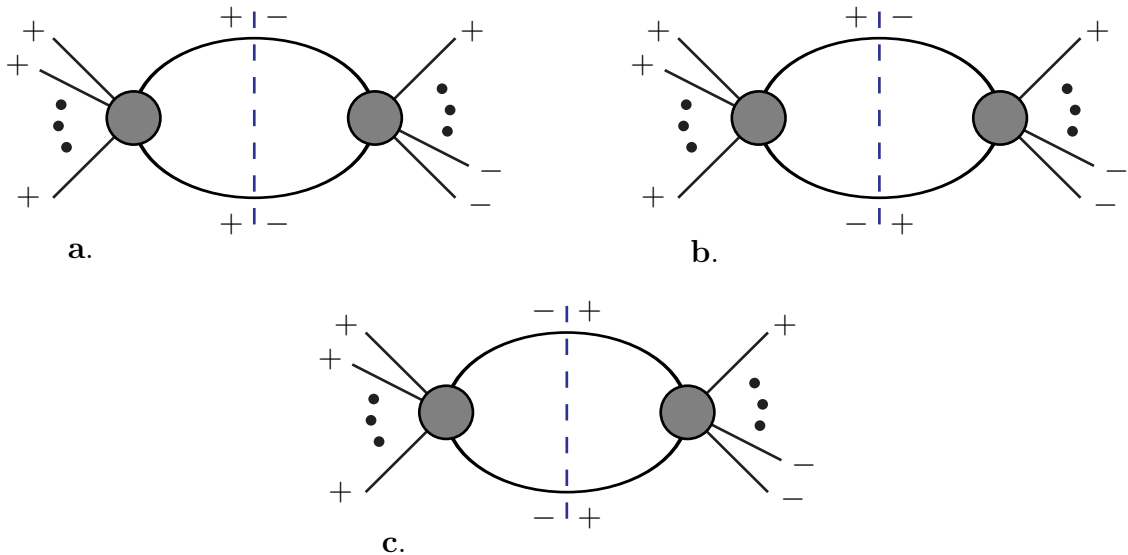

In the case where purely gluonic states are
crossing a cut, we consider the fully cut MHV amplitude. 
Examples of the helicity configurations corresponding to
the cuts, and thus the 
tree amplitudes that compose them, are 
depicted in figure~\ref{fig:gcut}:
the types of cut possible are 
\begin{itemize}
	\item[\textbf{a}.]
		 AP$\times {\rm N}^2$MHV
	 \item[\textbf{b}.]
		 UHV$\times$NMHV 
	 \item[\textbf{c}.]
		 MHV$\times$MHV.
\end{itemize}

\textbf{a}. These cuts do not contribute as the all-plus tree 
amplitude vanishes in any $p_\chi$ gauge~\cite{Boels:2011zz} 
\begin{align}
	A^{++;{\rm p_\chi}}(\mi{\bm l}_r^1,r^+,...,
			...,(s-1)^+,{\bm l}_{s}^1)
			= 0 \quad .
\end{align}

\textbf{b}. The UHV$\times$NMHV
cut can be extracted from the MHV-band amplitude in equation~
\bref{eq:mhvband} in $\gg\cdot q_{rs} = 0 $  
gauge~\cite{Craig:2011ws}
\begin{align}
	{\delta^\chi_{12}\delta^\chi_{34}\over 
	\mu^2\la \lambda_{l_r}
	\lambda_{l_s}\ra^2 }
	A(\mi{\bm l}_r^0,r^+,...,(s-1)^+,{\bm l}_{s}^0)
\end{align}
with 
\begin{align}
	\delta^\chi_{12} = 
	\biggl[	\delta^{(4)}\left( \la \lambda_i| \eta_{ia}\right)+
	{\mu\la \lambda_{l_r}\lambda_{l_r}\ra 
	\over \la \chi \lambda_{l_s}\ra
	\la \chi \lambda_{l_r} \ra} \delta^{(2)}
	\left( \la \chi \lambda_i\ra \eta_{ia}\right)\biggr] 
	\times \biggl[ 1- {[\lambda_{l_s}\chi ][\lambda_{l_r} 
	\chi]\over
	\mu[\lambda_{l_r}\lambda_{l_s}]}\delta^{(2)}
	\left({\mu\eta_{ia}\over
	[\lambda_i \chi ]}\right)
	\biggr]
	\label{eq:sde}
\end{align}
with the index $a\in \lbrace 1,2\rbrace$ in this case, 
and $i\in\lbrace \mi l_r,r,...,s-1,l_s\rbrace$.
$\delta^\chi_{34}$ takes the same form but carries the index
of the other $SU(2)$ branch of the broken $SU(4)$ 
$R$-symmetry~\cite{Craig:2011ws}.
We can apply the functional derivative
${\delta^4 \over \delta \eta_1^4}$ to extract the UHV case:
 the negatively spinning term 
comes entirely from the second term in the first bracket
giving
\begin{align}
	\delta^\chi_{12}\delta^\chi_{34}\rightarrow 
	{\mu^2\la \lambda_r\lambda_s \ra^2 }
	{\la \chi \lambda_{l_{r}}\ra^2\over \la \chi\lambda_{l_s}
	\ra^2} 
\end{align}
as is consistent with equation~\bref{eq:ap2v}.
The NMHV amplitude 
$A^{-+}({\bm l}_s^1,s,...,r-1,{\bm l}_r^1)$ 
comes from the term
\begin{align}
	\delta^\chi_{12}\delta^\chi_{34}\rightarrow 
	\delta^{(8)}\left( \la \lambda_i | \eta_{ia}\right)
	{\mu^2\over[\lambda_{l_r}\lambda_{l_s}]^2 }
	{[\lambda_{l_r}\chi ]^2\over [\lambda_{l_s}\chi]^2}\; ,
\end{align}
where in this case $i\in\lbrace s,s+1,...,r-1\rbrace$.
The gauge choice $p_\chi\cdot q_{rs}=0$ implies
\begin{align}
	[\chi|l_r|\chi \ra &=  [\chi| l_s|\chi\ra 
	\notag \\  
	= [\chi\lambda_{l_r}]\la \lambda_{l_r}\chi \ra
	&= [\chi\lambda_{l_s}]\la \lambda_{l_s}\chi \ra \,,
\end{align}
and thus
\begin{align}
	[\lambda_{l_r}\lambda_{l_s}]\la \lambda_{l_r}\lambda_{l_s}
	\ra
	= q_{rs}^2\,,
\end{align}
so that the generic UHV$\times$NMHV cut can be reduced to
\begin{align}
	&A^{-+}_L(\mi{\bm l}_r^1,r^+,...,
		...,(s-1)^+,{\bm l}_{s}^1)
		\times
	A^{+-}_R(\mi{\bm l}_s^1,s,...,r-1,{\bm l}_r^1)
	=
\notag \\ 
	& 		A_L(\mi{\bm l}_r^0,r^+,...,
		...,(s-1)^+,{\bm l}_{s}^0)
		\times
		{\la \chi \lambda_{l_r}\ra^2 [\lambda_{l_s}\chi]^2
		\over
		\la \chi \lambda_{l_s}\ra^2 [\lambda_{l_r}\chi]^2
		}
		\times
	{\delta^{(8)}\left( \la \lambda_i | \eta_{ia}\right)
	\mu^2
	\over q_{rs}^4}
	A_R(\mi {\bm l}_s^0,s^+,...,(r-1)^+,{\bm l}_r^0) \; .
	\label{eq:ttf}
\end{align}
After making the identification from equation~\bref{eq:mhvband}
(with corresponding arguments) and the definition in equation~\bref{eq:sde}
then we can match
\begin{align}
	{\delta^4\over \delta^4\eta_i}{\delta^4\over \delta^4\eta_j}
A_{\rm tree}^{\rm MHV-band}  = 
	{\delta^4\over \delta^4\eta_i}{\delta^4\over \delta^4\eta_m}
	{\delta^{(8)}\left( \la \lambda_i | \eta_{ia}\right)
	\mu^2
	\over q_{rs}^4}
	A_R({\bm l}_s^0,s^+,...,(r-1)^+,{\bm l}_r^0) \; ,
\end{align}
and we identify the right-most factor in equation~\bref{eq:ttf}
as the NMHV two-massive-scalar amplitude.
Thus, as 
\begin{align}
		{\la \chi (\mi\lambda_{l_r})\ra [\lambda_{l_r}\chi]
		\over
		\la \chi \lambda_{l_s}\ra [(\mi \lambda_{l_s})\chi]
		}=-1\, ,
\end{align}
		we get
\begin{align}
	&A^{-+}_L
		\times
	A^{+-}_R
	= 
	A^{00}_L
		\times
	A^{00}_R
	=A^{[0]}\biggr|^{\mu^2\neq0}_{q_{rs} \; {\rm cut}}
	\; .
\end{align}
From this, we identify the NMHV$\times$UHV contribution to
the $q_{rs}$ gluon cut channel as that originating from
$\mathcal{S}$ in the decomposition~\bref{eq:sdec}.
The $A^{+-}_L\times A^{-+}_R$ term
is the same to make the total contribution 
$2\mathcal{S}$.
Similar reasoning can be used to deduce a similar 
result for 
fermions
\begin{align}
	A^{\left[{1\over 2}\right]}
	\biggr|^{\mu^2\neq0}_{{\rm UHV}\times{\rm NMHV} 
	\; q_{rs} \; {\rm cut}}
	=-2A^{[0]}\biggr|^{\mu^2\neq0}_{q_{rs} \; {\rm cut}} \; ,
\end{align}
and their contributions sum up to cancel with the 
scalar states in the $\mathcal{N}=4$ super-multiplet. 
\begin{align}
	A_n^{MHV}\biggr|^{\mu^2\neq0}_{{\rm UHV}\times{ \rm NMHV} 
	\; q_{rs} \; {\rm cut}}
	= 0 \quad .
\end{align}
Of course this could all be done formally using the CSW rules
developed in~\cite{Kiermaier:2011cr,Elvang:2011ub},
but treating the particles explicitly shows 
how this gauge separates the structure in 
terms of the decomposition in equation~\bref{eq:sdec}. 

\textbf{c}.
This leaves the MHV$\times$MHV
cuts. 
These simply correspond to the complete unitarity cut
version of equation~\bref{eq:mhvcut}
\begin{align}
	\int &d^4\eta_{l_r}d^4\eta_{l_s}
		A^{\rm MHV\; tree}_L(\mi{\bm l}_r^1,r,
		...,(s-1),{\bm l}_{s}^1)
		\times
	A^{\rm MHV\; tree}_R(\mi{\bm l}_r^1,r,...,s-1,{\bm l}_s^1)
	=
	\notag \\
	&\int d^4\eta_{l_r}d^4\eta_{l_s}
		{\delta^{(8)}\left(L\right)\over \mu^2 \la 
		\lambda_{l_s}\lambda_{l_r}\ra^2}
		A_L(\mi{\bm l}_r^0,r,
		...,(s-1),{\bm l}_{s}^0)
		\times
	{\delta^{(8)}\left(R\right)\over \mu^2 \la 
		\lambda_{l_s}\lambda_{l_r}\ra^2}
	A_R(\mi{\bm l}_r^0,s,...,r-1,{\bm l}_r^0) \; ,
	\notag \\
	&\hspace{0.5in}L\equiv|i \ra \eta_{iA}, \;i\in \lbrace \lambda_{\mi l_r},
		r,...,	s-1,\lambda_{l_s}\rbrace
		\; ; \;
	R\equiv	|i	 \ra 
		\eta_{jA} ,\; i\in \lbrace \lambda_{l_s},
		r,...,
		s-1,\lambda_{\mi l_r}\rbrace\; ;
\end{align}
so the Grassman integration gives
\begin{align}
	A^{MHV}\biggr|^{\mu^2\neq 0}_{q_rs\;{\rm cut}}=&
	{\delta^{(8)}\left( \la \lambda_i | \eta_{iA}\right)
	\over \mu^4}A_L(\mi{\bm l}_r^0,r^+,...,(s-1)^+,{\bm l}_s^0)
	A_R(\mi{\bm l}_s^1,s^+,...,(r-1)^+,{\bm l}_r^1) \; ,
\end{align}
equivalent to equation~\bref{eq:cutproof2} upon applying the 
functional derivatives 
${\delta^4\over\delta\eta_i^4} 
	{\delta^4\over\delta\eta_j^4}$, 
\begin{align}
	A_n^{\mathcal{N}=4}(1^+,2^+,...,i^-,...,j^-,...,n^+)
	\biggr|^{\mu^2}_{q_{rs}\;{\rm cut}} 
	&={\delta^4\over\delta\eta_i^4} 
	{\delta^4\over\delta\eta_j^4}\left[ A^{\rm MHV}_n
	\biggr|^{\mu^2\neq 0}_{q_{rs}\;{\rm cut}}  
	\right]
\notag \\
	&= {\la ij \ra ^4\over 2\mu^4}A_n^{\rm AP}
	\biggr|_{q_{rs}\;{\rm cut}} 
\end{align}
thus proving the conjecture~\bref{eq:conj}.
\newpage

%% file: cuts-sec.tex

\section{Generalised cuts and all-epsilon forms}
\label{sec:gencut}
To compute the one-loop scattering amplitudes of the verified
conjecture~\bref{eq:conj} we present a new generalised version
of $D$-dimensional unitarity which extracts the coefficients
of a fixed basis of integrals.
The simplicity of one-loop amplitudes in general, 
and the all-plus amplitude in 
particular leads us to an all-multiplicity form to all orders in epsilon
which, through what is now the shift identity~\bref{eq:conj},
delivers the all-multiplicity form of the one-loop MHV $\mathcal{N}=4$
without any further computation.
The only master integrals necessary for these configurations
are pentagons and boxes, 
both of which have known closed forms in general dimension.

The basis we use was proposed 
by Giele, Kunszt, and Melnikov~\cite{Giele:2008ve}, 
and later used for the all-plus amplitude and other examples by 
Badger~\cite{Badger:2008cm}; we re-derive it in 
section~\ref{sec:basis} for completeness.
Techniques for fixing one-loop amplitudes by cuts
in four-dimensions generalise to provide direct 
techniques to extract the coefficients of the 
scalar integrals forming the smaller basis up 
to $\mathcal{O}(\epsilon)$ and 
rational terms.
In sections~\ref{sec:cutbox} and~\ref{sec:cutpent}
we show how we can use an equivalent technique
for the box and pentagon coefficients. 
Moreover, in the case of the box we see that the 
massive solutions can easily be parametrised in terms of
the four-dimensional null solutions.

\subsection{Integral bases}
\label{sec:basis}
The well-known principle of reducing loop-level amplitudes
to kinematic coefficients over a basis of known master-integrals
takes a simple form at one-loop. When truncating
to $\mathcal{O}(\epsilon)$, an amplitude $A_n$ can be 
expressed
\begin{align}
	A_n = \mathbf{d}_4\cdot \mathbf{I}^{D=4-2\epsilon}_4 
	+ \mathbf{d_3}\cdot\mathbf{I}^{D=4-2\epsilon}_3
	+\mathbf{d_2}\cdot\mathbf{I}^{D=4-2\epsilon}_2
	+d_R +\mathcal{O}(\epsilon)\quad .
	\label{eq:bas1}
\end{align}
This is a reduced version of 
a complete basis~\cite{Giele:2008ve,Ellis:2008ir,Badger:2008cm}
which includes pentagons, tadpoles and higher-dimensional scalar 
integrals
\begin{align}
	A_n = \mathbf{b}_5\cdot \mathbf{I}^{D=\bar{D}-2\epsilon}_5 
	+&\mathbf{b}_4(\epsilon)\cdot \mathbf{I}^{D=\lbrace 4,6,8\rbrace -2\epsilon}_4 
	+ \mathbf{b_3}(\epsilon)\cdot\mathbf{I}^{D=\lbrace 4,6\rbrace -2\epsilon}_3
	\notag \\
	+&\mathbf{b_2}(\epsilon)\cdot\mathbf{I}^{D=\lbrace 4,6\rbrace -2\epsilon}_2
	+\mathbf{b_1}(\epsilon)\cdot\mathbf{I}^{D=4-2\epsilon}_1 \quad.
	\label{eq:bas2}
\end{align}
We leave $\bar{D}$ as a general integer placeholder to be fixed later in
the derivation.
The epsilon dependence of the coefficient in fact takes a very
particular form, 
such that it integrates well with massive unitarity
cuts.
For completeness it is worth reviewing how this basis is 
derived~\cite{Badger:2008cm}.

The principles underlying the building of a basis
are both to
ensure that the basis is compatible
with the generalised unitarity technique which builds the
amplitude, in the sense that a given cut is precisely a 
prefactor of a given integral without overlaps or other
ambiguities,
and that it is general enough to capture all high-multiplicity
behaviour.
The most general basis one could write down includes all possible
one-loop $4-2\epsilon$ Feynman integrals with general 
numerator dependence on loop momentum
$\ell$, as emerges naturally
from a dimensionally regulated 
Feynman diagram construction in  four dimensions:
\begin{align}
	\left\lbrace I^{4\mi 2\epsilon}_m[\mathcal{P(\ell )}]
	\right\rbrace \quad .
\end{align}

\begin{figure}[ht]
\centerline{
    \begin{picture}(90,90)(-20,-220)    
	    \BCirc(0,-150){30}
	    \Line(-30,-150)(-60,-150)
	    \Line(30,-150)(60,-150)
	    \Vertex(-30,-150){2}
	    \Vertex(30,-150){2}
	    \Line(-80,-170)(-60,-150)
	    \Line(80,-130)(60,-150)
	    \Line(80,-170)(60,-150)
	    \Line(-80,-130)(-60,-150)
	    \GOval(-60,-150)(10,5)(0){0.5}
	    \GOval(60,-150)(10,5)(0){0.5}
	    \Text(0,-150)[c]{$\Huge{\leq \mu^{2}}$}
    \end{picture}
    \begin{picture}(150,150)(-120,-80)    
    \Line(-30,50)(-50,30)
    \Line(-30,-50)(-50,-30)
	    \Line(-60,0)(-50,30)
    \Line(-60,0)(-50,-30)
    \Line(30,50)(50,30)
    \Line(30,-50)(50,-30)
	    \Line(60,0)(50,30)
    \Line(60,0)(50,-30)
	\DashLine(30,50)(-30,50){2}
	\DashLine(30,-50)(-30,-50){2}
	    \Vertex(-30,50){2}
	    \Vertex(-50,30){2}
	    \Vertex(-30,-50){2}
	    \Vertex(-50,-30){2}
	    \Vertex(-60,0){2}
	    \Vertex(60,0){2}
	    \Vertex(50,30){2}
	    \Vertex(50,-30){2}
	    \Vertex(30,-50){2}
	    \Vertex(30,50){2}
    \Line(-30,50)(-45,70)
    \Line(-70,45)(-50,30)
    \Line(30,-50)(45,-70)
    \Line(70,-45)(50,-30)
    \Line(-30,-50)(-45,-70)
    \Line(-70,-45)(-50,-30)
    \Line(60,0)(85,0)
    \Line(-60,0)(-85,0)
    \Line(30,50)(45,70)
    \Line(70,45)(50,30)
	    \Text(86,8)[c]{$p_r$}
	    \Text(-86,8)[c]{$p_1$}
	    \Text(0,0)[c]{$\Huge{\leq \mu^{2\left\lfloor {n\over
	    2}\right\rfloor}}$}
    \end{picture} 
    }
	\caption[UV limits on bubble]{Ultraviolet limits on 
	the bubble integral. 
	Left is the possible sources of $\mu^{2r}$ numerators in 
	renormalisable gauge theories, each
	vertex contributing a possible $\ell^\mu$ to the numerator.
	The only possible contributions where $r>1$ comes from 
	higher point integrals, which we include separately in the
	basis.}
    \label{fig:masint}
\end{figure}
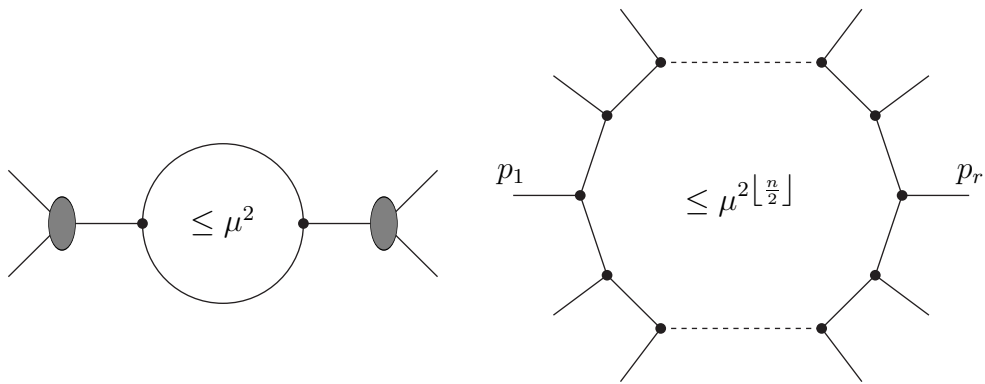

There are further constraints we can apply to this
polynomial.
The theory we are working in is renormalisable. 
This puts a limit on the mass dimension of the vertices,
and thus through power counting on the order of the polynomial
$\mathcal{P}$ for a given one-loop graph.
For example the bubble Feynman diagram depicted in 
figure~\ref{fig:masint}
 could only
contribute numerators at worst quadratic in the
loop momentum. 

It is also very well known that basis elements possessing 
numerator terms 
involving the four-dimensional components of the loop momentum
$l^\mu$ can be expressed in terms of scalar 
integrals~\cite{Passarino1979,Neerven1984,Bern:1993kr}.
On any given 
cut, this implies that the 
numerator $\mathcal{P}$ becomes a polynomial in
the ``mass" term $\mu^2$ from the $(-2\epsilon)$-dimensional component of 
the momentum (see section~\ref{sec:cuts}).
We choose to represent numerators in this form, which allows us to conclude that~\cite{Bern:1995db}
\begin{align}
	I^{4 -  2\epsilon}_m[\mu^{2r}] = 
	-\epsilon(1-\epsilon)\cdots(r-1-\epsilon)
	I^{4+2r -  2\epsilon}_m\, .
	\label{eq:mue}
\end{align}
Although $r\leq \left\lfloor {n\over 2} \right\rfloor $
for diagrams of the type in figure~\ref{fig:masint},
there is the question of how the power counting
follows the reduction
of ($m\geq 6)$-point integrals to pentagons. 
Moreover, using generalised unitarity to fix the coefficient
of the pentagon integral functions leads to an ambiguity as to
the dimension they are defined in.

This can be understood from the fact that the $\epsilon$-dependence
of the coefficients in the
integral dimension-shift identity~\cite{Bern:1993kr}
\begin{align}
	I_5^{D}= {1\over 2} 
	\left[\sum_{j=1}^n c_jI_4^{D;(j)}
	+(4-D) c_0I^{D+2}_5\right]
	\label{eq:pentred}\; ,
\end{align}
is identical to that arising from a $\mu^2$ being inserted into the numerator
of $I_5^{D}$.
Applying the identity~\bref{eq:pentred} into itself $r$ 
times gives
\begin{align}
	I_5^{4 - 2\epsilon} &= 
	-\epsilon(1-\epsilon)\cdots(r-1-\epsilon)
	I_5^{^{4+2r -  2\epsilon}} + {\rm boxes}
	\notag \\
&= 
	I_5^{^{4 - 2\epsilon}}[\mu^{2r}] + {\rm boxes}
\end{align}
so that the choice $r$ in this case is determined by the 
choice of dimension-shifted 
boxes that are included in the basis.
Choosing to include
boxes $ I_4[\mu^{2r}]$ implies
that we can restrict ourselves to pentagon coefficients of $I_5[\mu^{2r+2}]$,
as the box cuts compute the contribution from the first term on 
the right-hand side of equation~\bref{eq:pentred} for 
that respective $r$. 
The choice of which boxes to include is thus contingent on the
choice of pentagon, and the cut computes the
independent contributions alongside
the contribution from a pentagon shifted from $0$ to 
$r+1$.
As could be deduced from the box version of 
figure~\ref{fig:masint}, there is an
upper limit of $r=2$ for a
given box to exist independently from higher point functions.
This implies the truncation
of polynomials in $\mu^2$ to quadratic order ($\mu^4$).

The external momenta are four-dimensional, 
and thus there is the additional simplification that 
$n$-point integrals reduce to pentagons for $n> 5$.
This fact 
can be seen in a number of ways~\cite{Bern:1993kr}, and recently 
a simple formula was written down by one of the 
authors~\cite{Jehu:2020xip} 
which generalises very simply for any $n$.
Crucially, there is a very simple coincidence of interpretation
with maximal unitarity cuts, which allows the coefficient of
the reduction to 
simply be interpreted as propagators frozen at a cut solution.
This compatibility permits us to interpret, for $m>5$,
\begin{align}
	I^{4\mi 2\epsilon}_m\biggr|_{[i_1,i_2,i_3,i_4,i_5]\;{\rm cut} }
	= \left[\prod_{j=1}^{m-5}\xi_j\right]I^{[i_1,i_2,i_3,i_4,i_4];4\mi 2\epsilon}_5 \; ,
\end{align}
so capturing the pentagons will also 
reproduce all higher-point
integrals, thus extending generalised unitarity 
to treat complete unitarity cuts.
We emphasise here that all degrees of freedom are fixed,
so unlike the box case $\mu^2$ is not a degree of freedom for 
the ($m\geq 5$)-point integrals. 
The general solution to the penta-cut is~\cite{Jehu:2020xip}
\begin{align}
	\mu^2\bigr|_{[i_1,i_2,i_3,i_4,i_5]\;{\rm cut} }
	&= {1\over c_0} = {16\Upsilon_5\over
				\Delta_5}
	\notag \\
	&=
			{\tr(q_{i_1i_2}q_{i_2i_3}
	q_{i_3i_4}q_{i_4i_5}q_{i_5i_1}q_{i_1i_2}q_{i_2i_3}
	q_{i_3i_4}q_{i_4i_5}q_{i_5i_1})- 2\prod_{k=1}^5
	q_{i_ki_{k+1}}^2\over
				\tr_5^2(q_{i_1i_2}q_{i_2i_3}
	q_{i_3i_4}q_{i_4i_5})}\quad .
\end{align}
Thus we can refine the basis~\bref{eq:bas2}, as we know what
the $\epsilon$-dependence of the coefficients must be, 
and so we bundle the dependence into the integral basis.
Knowing that boxes can arise independently for 
$I_m[\mu^{2r}]$,
we also fix the dimension of the pentagon to $D=10-2\epsilon$ (i.e.\ $r=3$)
such that boxes are included up to $I_4[\mu^4]$ separately in the basis.
For massless theories we drop tadpoles; our final basis is thus
\begin{align}
	A_n = \mathbf{b}_5\cdot \mathbf{I}_5[\mu^6] 
	+\mathbf{b}_4\cdot \mathbf{I}_4\left[
1,\mu^2,\mu^4
	\right]
	&+ \mathbf{b}_3\cdot\mathbf{I}_3\left[
1,\mu^2	\right]
	+\mathbf{b}_2\cdot\mathbf{I}_2
	\left[
	1,\mu^2
	\right]
	\quad .
	\label{eq:bas3} 
\end{align}
The ingredients needed to construct the 
coefficients are the tree amplitudes in
equation~\bref{eq:ap2s} which we use 
to compute the one-loop amplitude 
$A_n^{\rm AP} = 2A_n^{[0]}$.
The bubbles and triangles vanish trivially from the scaling of the tree amplitudes

\begin{align}
A^{\rm tree}&(\mathbf{1}^0,
	2^+,3^+,...,(n-1)^+,\mathbf{n}^0)
	\sim \mu^2
\end{align}
so that a nonvanishing product of tree amplitudes would contribute only 
to order $\mathcal{O}(\mu^4)$, which is entirely captured by the box
integrals, by the above arguments. In other words, we can set $\mathbf{b}_3=\mathbf{b}_2=0$ in equation~\bref{eq:bas3}. We will proceed to show that the coefficients of $\mathbf{I}_4\left[\mu^4	\right]$ vanish as well.

\subsection{The cut box}
\label{sec:cutbox}
To compute the box coefficients which compose 
${ \rm \mathbf{b}}^{\rm AP}_4 $ we use a method 
analogous to how generalised
four-dimensional cuts can be used to determine the coefficient
of the $\mathcal{N}=4$ MHV amplitude.

\begin{figure}[ht]
\centerline{
    \begin{picture}(140,140)(-50,-34)    
	    \SetWidth{1}
     \Line( 0, 0)( 0,60)
     \Line( 0,60)(60,60)
     \Line(60,60)(60, 0)
     \Line(60, 0)( 0, 0)
     \Line( 0, 0)(-20,-20)
     \Line(60,60)(80,80)
     \Line(0,60)(-20,80)
     \Line(60, 0)(80,-20)
   \CCirc(60,60){7}{Black}{Gray} 
   \CCirc(0,60){7}{Black}{Gray} 
   \CCirc(60,0){7}{Black}{Gray} 
   \CCirc(0,0){7}{Black}{Gray} 
    \SetColor{Blue}
     \DashLine(30,-10)(30,10){5}
     \DashLine(30,50)(30,70){5}
     \DashLine(-10,30)(10,30){5}     
     \DashLine(50,30)(70,30){5}
    \end{picture} 
    \begin{picture}(140,100)(-50,-34)    
	    \SetWidth{1}
     \Line( 0, 0)( 0,60)
     \Line( 0,60)(60,60)
     \Line(60,60)(60, 0)
     \Line(60, 0)( 0, 0)
     \Line( 0, 0)(-20,0)
     \Line( 0, 0)(0,-20)
     \Line(60,60)(80,80)
     \Line(0,60)(-20,80)
     \Line(60, 0)(80,-20)
   \CCirc(60,60){7}{Black}{Gray} 
   \CCirc(0,60){7}{Black}{Gray} 
   \CCirc(60,0){7}{Black}{Gray} 
   \CCirc(0,0){7}{Black}{Gray} 
	    \Vertex(-12,-7){2}
	    \Vertex(-7,-12){2}
    \SetColor{Blue}
     \DashLine(30,-10)(30,10){5}
     \DashLine(30,50)(30,70){5}
     \DashLine(-10,30)(10,30){5}     
     \DashLine(50,30)(70,30){5}
    \end{picture} 
    \begin{picture}(140,100)(-50,-34)    
	    \SetWidth{1}
     \Line( 0, 0)( 0,60)
	    \LongArrow(-15,15)(-15,45)
	    \Text(-25,30)[c]{$l_{1}$}
     \Line( 0,60)(60,60)
     \Line(60,60)(60, 0)
     \Line(60, 0)( 0, 0)
     \Line( 0, 0)(-20,-20)
     \Line(60,60)(80,80)
     \Line(0,60)(0,80)
     \Line(0,60)(-20,60)
     \Line(60, 0)(80,0)
     \Line(60, 0)(60,-20)
   \CCirc(60,60){7}{Black}{Gray} 
   \CCirc(0,60){7}{Black}{Gray} 
   \CCirc(60,0){7}{Black}{Gray} 
   \CCirc(0,0){7}{Black}{Gray} 
     \Text(-8,-25)[r]{${i_4}$}   
     \Text(-20,60)[r]{${i_1}$}   
     \Text(70,87)[l]{${i_2}$}  
     \Text(80,0)[l]{${i_3}$}   
	    \Vertex(72,-7){2}
	    \Vertex(67,-12){2}
	    \Vertex(-12,67){2}
	    \Vertex(-7,72){2}
    \SetColor{Blue}
     \DashLine(30,-10)(30,10){5}
     \DashLine(30,50)(30,70){5}
     \DashLine(-10,30)(10,30){5}     
     \DashLine(50,30)(70,30){5}
    \end{picture} 
    \begin{picture}(50,50)(-25,-25)    
	    \Text(-10,40)[c]{$\sim \mu^4$}
    \end{picture} 
   } 
\centerline{
    \begin{picture}(140,100)(-50,-20)    
	    \SetWidth{1}
     \Line( 0, 0)( 0,60)
     \Line( 0,60)(60,60)
     \Line(60,60)(60, 0)
     \Line(60, 0)( 0, 0)
     \Line( 0, 0)(-20,0)
     \Line( 0, 0)(0,-20)
     \Line(60,60)(80,80)
     \Line(0,60)(0,80)
     \Line(0,60)(-20,60)
     \Line(60, 0)(80,-20)
   \CCirc(60,60){7}{Black}{Gray} 
   \CCirc(0,60){7}{Black}{Gray} 
   \CCirc(60,0){7}{Black}{Gray} 
   \CCirc(0,0){7}{Black}{Gray} 
	    \Vertex(-12,-7){2}
	    \Vertex(-7,-12){2}
	    \Vertex(-12,67){2}
	    \Vertex(-7,72){2}
    \SetColor{Blue}
     \DashLine(30,-10)(30,10){5}
     \DashLine(30,50)(30,70){5}
     \DashLine(-10,30)(10,30){5}     
     \DashLine(50,30)(70,30){5}
    \end{picture} 
    \begin{picture}(140,100)(-50,-20)    
	    \SetWidth{1}
     \Line( 0, 0)( 0,60)
     \Line( 0,60)(60,60)
     \Line(60,60)(60, 0)
     \Line(60, 0)( 0, 0)
     \Line( 0, 0)(-20,0)
     \Line( 0, 0)(0,-20)
     \Line(60,60)(80,80)
     \Line(0,60)(0,80)
     \Line(0,60)(-20,60)
     \Line(60, 0)(80,0)
     \Line(60, 0)(60,-20)
   \CCirc(60,60){7}{Black}{Gray} 
   \CCirc(0,60){7}{Black}{Gray} 
   \CCirc(60,0){7}{Black}{Gray} 
   \CCirc(0,0){7}{Black}{Gray} 
	    \Vertex(-12,-7){2}
	    \Vertex(-7,-12){2}
	    \Vertex(72,-7){2}
	    \Vertex(67,-12){2}
	    \Vertex(-12,67){2}
	    \Vertex(-7,72){2}
    \SetColor{Blue}
     \DashLine(30,-10)(30,10){5}
     \DashLine(30,50)(30,70){5}
     \DashLine(-10,30)(10,30){5}     
     \DashLine(50,30)(70,30){5}
    \end{picture} 
    \begin{picture}(140,100)(-50,-20)    
	    \SetWidth{1}
     \Line( 0, 0)( 0,60)
     \Line( 0,60)(60,60)
     \Line(60,60)(60, 0)
     \Line(60, 0)( 0, 0)
     \Line( 0, 0)(-20,0)
     \Line( 0, 0)(0,-20)
     \Line(60,60)(60,80)
     \Line(60,60)(80,60)
     \Line(0,60)(0,80)
     \Line(0,60)(-20,60)
     \Line(60, 0)(80,0)
     \Line(60, 0)(60,-20)
   \CCirc(60,60){7}{Black}{Gray} 
   \CCirc(0,60){7}{Black}{Gray} 
   \CCirc(60,0){7}{Black}{Gray} 
   \CCirc(0,0){7}{Black}{Gray} 
	    \Vertex(-12,-7){2}
	    \Vertex(-7,-12){2}
	    \Vertex(72,-7){2}
	    \Vertex(67,-12){2}
	    \Vertex(-12,67){2}
	    \Vertex(-7,72){2}
	    \Vertex(72,67){2}
	    \Vertex(67,72){2}
    \SetColor{Blue}
     \DashLine(30,-10)(30,10){5}
     \DashLine(30,50)(30,70){5}
     \DashLine(-10,30)(10,30){5}     
     \DashLine(50,30)(70,30){5}
    \end{picture} 
    \begin{picture}(50,50)(-25,-25)    
	    \Text(-10,40)[c]{$\sim\mathcal{O}(\mu^6)$}
    \end{picture} 
   } 
	\caption[Box Cuts]{TOP: Diagrams which contribute to the basis~\bref{eq:bas3}.
	BOTTOM: Diagrams which are eliminated by the truncation of terms $\mathcal{O}(\mu^6)$.}
    \label{fig:cutbox}
\end{figure}
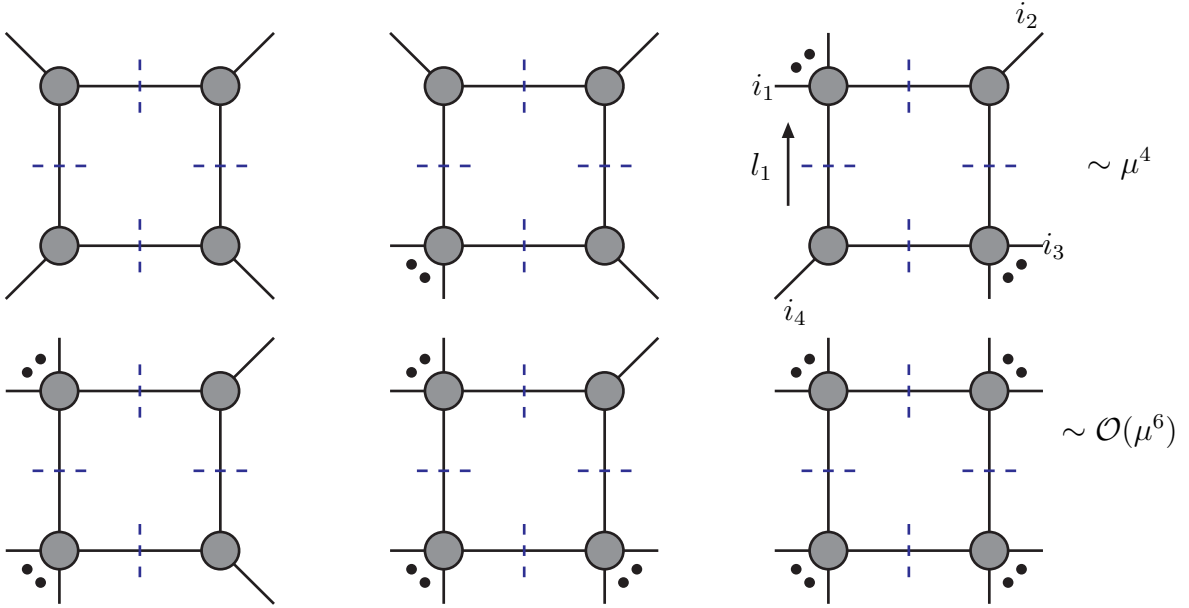

The four-dimensional cut conditions for massless propagators
\begin{align}
	\ell_{i_k}^2 = l_{i_k}^2 = 0 \;, 
	\quad k \in \lbrace 1,2,3,4\rbrace
	\label{eq:ccc}
\end{align}
are readily solved.  
For the massless, one-mass and two-mass boxes 
depicted in the top row 
in figure~\ref{fig:cutbox} the solution to the 
conditions~\bref{eq:ccc} is 
\begin{align}
	l^\mu_{i_1}=\bar{l}^\mu_{\pm} = 
	{\tr_\pm(q_{i_1i_2}i_2i_4\gamma^\mu)\over  s_{i_2i_4}}
	\label{eq:mlsol}
\end{align}
which can also be expressed in spinor-helicity formalism as
\begin{align}
	(\bar{l}_+)_{a\dot{b}} &= \left({|i_4][i_2|q_{i_1i_2} \over
				[i_2i_4]}\right)_{a\dot{b}}	
				\; ,
				\notag \\
	(\bar{l}_-)_{a\dot{b}} &= \left({q_{i_1i_2}|i_2\ra \la i_4| \over
				\la i_4i_2\ra }\right)_{a\dot{b}}	
				\quad .
\end{align}
For massive propagators, the cut conditions are
\begin{align}
	\ell_{i_k}^2 = l_{i_k}^2 - \mu^2 = 0 \;,\quad 
	k \in \lbrace 1,2,3,4\rbrace\,,
	\label{eq:cdx}
\end{align}
which have an extra unfixed parameter compared to the massless cut
conditions in equation~\bref{eq:ccc}. 
Intuitively, to match up with the basis~\bref{eq:bas3}, 
this parameter can simply be $\mu^2$, 
however a more generally applicable
solution is to choose it to be a dimensionless 
parameter $\alpha$,
such that we can define the solution to equations~\bref{eq:cdx} in
terms of the massless solutions~\bref{eq:mlsol}:
\begin{equation}
	l_1^\mu=(\alpha \bar{l}_++(1-\alpha )\bar{l}_-)^\mu 
	\; ,
	\label{eq:all}
\end{equation}
with the relationship between $\alpha$ and $\mu$ defined by 
\begin{align}
	\alpha(1-\alpha) &= { \mu^2\over 2l_+\cdot l_-}\quad .
	\label{eq:qua}
\end{align}
It is convenient to note that 
$2l_+\cdot l_-= (c^{\rm box}_0)^{- 1}$ for 
the following representation\footnote{The box coefficient can also be expressed as a ratio of 
kinematic determinants~\cite{Melrose1965,Bern:1993kr,Jehu:2020xip}.}
of $c^{\rm box}_0$,
\begin{align}
	c^{\rm box}_0 = 
	{4 s_{i_2i_4}\over 
	[i_2|q_{i_1i_2}|i_2\ra [i_4|q_{i_1i_2}|i_4\ra
	}\;.
\end{align}
We can then write solutions to equation~\bref{eq:qua}
as
\begin{align}
	\alpha_\pm &=  {1\pm \sqrt{1+c^{\rm box}_0
	\mu^2}\over 2} \quad .
\end{align}
The 
solution to~\bref{eq:cdx} is  split into two branches
which meet at $\alpha={1\over 2}$.
The two solutions simply correspond to the choice 
of parametrisation, as it is 
trivial to see that the replacement
\begin{align}
	\alpha &\rightarrow 1-\alpha\; , \; \pm \rightarrow \mp
\end{align}
leaves~\bref{eq:all} unaltered.
When making the choice, one 
needs to average over the solutions, 
just like for the massless case.
Instead of making the solution depend on choice of $\alpha_\pm$ we 
fix $\alpha\equiv \alpha_+$ then our two solutions can be 
expressed
\begin{align}
	l_\pm = \alpha \bar{l}_\pm + (1-\alpha)\bar{l}_\mp
\end{align}
These solutions are then 
input into the massive cuts of the kind 
depicted in figure~\ref{fig:cutbox}.

As the three-point amplitude with one
on-shell gluon and two equal-mass scalars is not 
defined by equation~\bref{eq:ap2s}, 
we define it here~\cite{Schwinn2005,Arkani-Hamed:2017jhn}:
\begin{align}
	A_3^{\rm tree}(i^+,{\bm l}_{i+1}^0,{\bm l}_i^0) 
	= {[i| l_i | \chi\ra\over \la i \chi \ra}\,.
	\label{eq:ap2s3}
\end{align}
We see that unlike the leading term of the 
four-point and above UHV amplitudes,
the amplitude~\bref{eq:ap2s3} does not scale
linearly with $\mu^2$.
As with the bubbles and triangles,
we will use the truncation of the $\mu^2$ polynomial 
to rule out two-mass hard, three-mass and four-mass boxes 
which form the bottom diagrams of figure~\ref{fig:cutbox} from
the functional form.
To this end, and to simplify future manipulations, we
explicitly consider some basic simplifications
of adjacent three-point amplitudes.
Without loss of generality considering the adjacent massive cuts 
$l_1^2=l_2^2=l_3^2=\mu^2$; we can express the corresponding product
of amplitudes thus
\begin{align}
	A_3^{\rm tree}(1^+,-{\bm l}^0_1,{\bm l}^0_2)
	\times
	A_3^{\rm tree}(2^+,-{\bm l}^0_2,{\bm l}^0_4)
	&= -{[1| l_1 | \chi\ra\over \la 1\chi \ra}
	\times {[2| l_1 | \chi\ra\over \la 2 \chi \ra}
	\notag \\
	&= {\la \chi|l_1 12l_1|\chi\ra\over \la 1\chi\ra \la 2\chi\ra \la 12\ra}
	\notag \\
	&= \mu^2 {[12]\over \la 12\ra} \,,
	\label{eq:3t4}
\end{align}
where we have made use of the fact that the cut conditions
imply $l_1\cdot p_1=l_1\cdot p_2 = 0$ to commute the $l_1$s together
in the last line using the identity
\begin{align}
	\la X|\lbrace a,b\rbrace |Y\ra = 2p_a \cdot p_b\la XY\ra \quad .
\end{align}

Equation~\bref{eq:3t4} implies that, like $n\geq 4$-point off-shell amplitudes,
adjacent three-point vertices have
an $\mathcal{O}(\mu^2)$ scaling, implying that two-mass hard boxes
have polynomials which start at $\mathcal{O}(\mu^6)$, like three-mass boxes; 
four-mass boxes begin at $\mathcal{O}(\mu^8)$. 
Recalling that our choice of basis in equation~\bref{eq:bas3} implies the truncation
of this polynomial up to $\mathcal{O}(\mu^4)$, we assert that,
much like the massless $\mathcal{N}=4$ cuts, 
the all-plus scalar cuts
only have contributions from two-mass-easy, 
one-mass and, in the four-point case,
massless boxes.

The coefficients of the boxes are 
captured by inputting the cut solutions given in
equation~\bref{eq:all} into the appropriate amplitudes,
averaging over $\alpha_{\pm}$ and truncating terms 
$\mathcal{O}(\mu^6)$.
Because of this truncation, the computation is greatly simplified,
as generically from the definition of amplitudes with 
two internal scalars~\bref{eq:ap2s},
\begin{align}
	A^{\rm tree}\left(\mi {\bm l}^0_{i_1},i_1^+,...,(i_2 - 1)^+
	,{\bm l}_{i_2}^0\right)
	= {\mu^2[i_1|l_{i_1}q_{i_1i_2}|i_2 -1]+ \mathcal{O}(\mu^4)\over 
	(s_{l_{i_1}i_1}-\mu^2)\la i_1(i_1+1) \cdots (i_2 - 2)
	(i_2- 1)\ra (s_{l_{i_2}(i_2-1)}-\mu^2)}\; ,
\end{align}
so we can truncate terms $\sim\mathcal{O}(\mu^4)$,
as they contribute to the cut to  $\mathcal{O}(\mu^6)$
at most. 
We double our expression of the coefficient 
$b^{[i_1,i_3 -  1,i_3,i_1 - 1]}_4$ of 
the $8-2\epsilon$ dimensional 
two-mass-easy box integral 
$I_4^{[i_1,i_3-  1,i_3,i_1-  1]}[\mu^4]$
as
$A^{AP}_n = 2A^{[0]}$ with the all-plus configuration
\begin{align}
	b_4^{[i_1,i_3 -  1,i_3,i_1 -  1]} &= 
	{1\over \mu^4} \left[2\times {1\over 2} \sum_{\alpha_\pm}
	A^{\rm tree}
	\times 
	A^{\rm tree}
\times
	A^{\rm tree}
	\times A^{\rm tree}
	\biggr|_{\mathcal{O}(\mu^6)}\right]
	\notag \\
	&= {1\over 2} {\tr(i_1q_{i_1i_3}i_3q_{i_3+1,i_1-1})\over
	\la 12 ...n1\ra }\; ,
	\label{eq:b4}
\end{align}
where $|_{\mathcal{O}(\mu^6)}$ denotes truncation of terms 
$\mathcal{O}(\mu^6)$; we suppress the arguments of the 
two-scalar all-plus tree amplitudes in the cut.
The result~\bref{eq:b4} 
matches the expected coefficient from the shifted version of the
$\mathcal{N}=4$ 
coefficient of the box given in equation~\bref{eq:n4trunc}.
\subsection{Pentagon coefficients}
\label{sec:cutpent}
We introduce the maximal cut unitarity solution to solve for the
pentagon coefficient. 
The solution is remarkably compact for all 
descendant pentagons~\cite{Jehu:2020xip}
\begin{figure}[ht]
\centerline{
	\begin{picture}(100,120)(-50,-60)    
	\Line(-30,-40)(-55,-46)
		\Vertex(-44,-49){2}
		\Vertex(-38,-55){2}
	\Line(-30,-40)(-35,-65)
	\Text(-28,-60)[l]{$i_1$}   
	\Line(30,-40)(55,-46)
		\Vertex(44,-49){2}
		\Vertex(38,-55){2}
	\Line(30,-40)(35,-65)
	\Text(59,-40)[r]{$i_5$}   
	\Line(-65,18)(-45,5)
	\Line(-65,-8)(-45,5)
		\Vertex(-60,0){2}
		\Vertex(-60,10){2}
	\Text(-64,-10)[r]{$i_2$}   
	\Line(65,18)(45,5)
	\Line(65,-8)(45,5)
		\Vertex(60,0){2}
		\Vertex(60,10){2}
	\Text(69,18)[l]{$i_3$}   
	\Line(14,60)(0,40)
	\Line(-14,60)(0,40)
	\Vertex(-5,57){2}
	\Vertex(5,57){2}
	\Text(-16,58)[r]{$i_4$}   
		\LongArrow(10,-53)(-10,-53)
		\Text(0,-60)[c]{$l$}
\SetWidth{2}
	\Line(-30,-40)(30,-40)
	\Line(45,5)(30,-40)
	\Line(-30,-40)(-45,5)
	\Line(0,40)(-45,5)
	\Line(45,5)(0,40)
\SetWidth{1}
   \CCirc(-30,-40){7}{Black}{Gray} 
   \CCirc(30,-40){7}{Black}{Gray} 
   \CCirc(45,5){7}{Black}{Gray} 
   \CCirc(-45,5){7}{Black}{Gray} 
   \CCirc(0,40){7}{Black}{Gray} 
	    \SetColor{Blue}
     \Line(0,-50)(0,-30)
     \Line(-50,-20)(-25,-11)
     \Line(50,-20)(25,-11)
     \Line(-30,34)(-11,14)
     \Line(30,34)(11,14)
    \end{picture} 
    }
	\caption[The penta cut]{The pentagon cut.}
    \label{fig:5cut}
\end{figure}
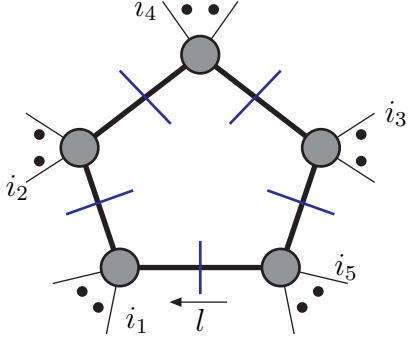    

The five-particle maximal cut depicted in 
figure~\ref{fig:5cut} fixes the loop momentum completely: 
\begin{align}
	l_{i_1}^\mu&=-{\tr_5\left(q_{i_1i_2}q_{i_2i_3}
	q_{i_3i_4}q_{i_4i_5}q_{i_5i_1}\gamma_\mu\right)\over 
			2\tr_5(q_{i_1i_2}q_{i_2i_3}
	q_{i_3i_4}q_{i_4i_5})}
	\notag \\
			\mu^2 = c_0^{-1} &= 
			{\tr(q_{i_1i_2}q_{i_2i_3}
	q_{i_3i_4}q_{i_4i_5}q_{i_5i_1}q_{i_1i_2}q_{i_2i_3}
	q_{i_3i_4}q_{i_4i_5}q_{i_5i_1})- 2\prod_{k=1}^5
	q_{i_ki_{k+1}}^2\over
				\tr_5^2(q_{i_1i_2}q_{i_2i_3}
	q_{i_3i_4}q_{i_4i_5})}
						\label{eq:5cutsol}
\end{align}
One now need simply draw all possible pentagons.
The coefficient computed from the product of amplitudes
is naturally interpreted as 
that of the four-dimensional pentagon
$I_5[1]$, but as we have already computed box coefficients up 
to $I_4[\mu^4]$, part of this function's contribution to the amplitude
has already been captured.
Applying the dimension-shift relation~\bref{eq:pentred} three times up to $I_5[\mu^6]$
introduces a factor $c_0^3$.
The coefficient of $I^{[i_1,i_2,i_3,i_4,i_5]}_5[\mu^6]$ is thus
\begin{align}
	b_5^{[i_1,i_2,i_3,i_4,i_5]} = c_0^3 
	A^{\rm tree}(\mi{\bm l}^0_{i_1}, i_1^+,...,(i_2- 1)^+,
	{\bm l}^0_{i_2})\times
	A^{\rm tree}(\mi{\bm l}^0_{i_2},i_2^+,...,(i_3- 1)^+,
	{\bm l}^0_{i_3})&\times
	\notag \\
	A^{\rm tree}(\mi{\bm l}^0_{i_3},i_3^+,...,(i_4- 1)^+,{\bm l}^0_{i_3})\times 
	A^{\rm tree}(\mi{\bm l}^0_{i_4},i_4^+,...,(i_5- 1)^+,{\bm l}^0_{i_3})&\times
	\notag \\
	A^{\rm tree}(\mi{\bm l}^0_{i_5},i_5^+,...,(i_1- 1)^+,{\bm l}^0_{i_3})&\quad .
	\label{eq:b5}
\end{align}
Note that propagator factors
can be substituted for the general $(n>5)$-multiplicity kinematic 
coefficients of the integral reduction 
$I_n\rightarrow I_5$~\cite{Jehu:2020xip}
\begin{align}
	{1\over q^2_{\mi l_{i_k}r}- \mu^2}\biggr|_{[i_1,i_2,i_3,i_4,i_5] 
			{\rm cut}} = {1\over 2 }\xi^{[i_1,i_2,i_3,i_4,i_5,r]}_r
			\; ,
			\label{eq:subs}
\end{align}
where 
\begin{align}
\xi^{[i_1,i_2,i_3,i_4,i_5,r]}_r
	={-2\tr_5(q_{i_1i_2}q_{i_2i_3}
	q_{i_3i_4}q_{i_4i_5})\over \tr_5(q_{i_1i_2}q_{i_2i_3}
	q_{i_3i_4}q_{i_4i_5}q_{i_5r}q_{ri_1})}\quad .
\end{align}
With the solution~\bref{eq:5cutsol}, the substitution~\bref{eq:subs}
and the explicit form of the off-shell scalar amplitude~\bref{eq:ap2s}, 
equation~\bref{eq:b5} is a closed form for any given pentagon coefficient.

\subsection{Six-point example}
We demonstrate the technique explicitly by computing the first non-trivial example:
the six-point amplitude
$A^{\rm AP}_6$.
There are two possible box configurations,
one mass and two-mass-easy. 
The box cuts are thus of the form
\begin{align}
	b_4^{[a,b,c,f]} &= {1\over \mu^4}\sum_{l_\pm}\left[ A^{\rm tree}_3\times A^{\rm tree}_3\times 
	A^{\rm tree}_5\times A^{\rm tree}_3 \right]\biggr|_{\mathcal{O}(\mu^6)}\; , 
	\notag \\
	b_4^{[a,c,d,f]} &= {1\over \mu^4}\sum_{l_\pm}\left[ A^{\rm tree}_4
	\times A^{\rm tree}_3\times 
	A^{\rm tree}_4\times A^{\rm tree}_3 \right]\biggr|_{\mathcal{O}(\mu^6)}\; .
\end{align}
We can simplify the computation by massaging these expressions before inputting 
the explicit solutions $l_{\pm}$ defined in equation~\bref{eq:all}.
Using the identity~\bref{eq:3t4}
\begin{align}
\	\mu^4 b_4^{[a,b,c,f]} &= \sum_{l_\pm}\left[ \mu^2 {[fa]\over \la fa \ra} \times 
	{\mu^2[c|\mi l_cq_{ce}|e][b|l_c|\chi \ra
	\over (s_{\mi l_cc}-\mu^2)\la cd \ra \la de \ra \la b\chi\ra (s_{el_f}-\mu^2)}\right]\biggr|_{\mathcal{O}(\mu^6)}\; , 
\end{align}
where the left-most factor is a combination of the two three-point
amplitudes with legs $f$ and $a$.
If we factor out the Parke-Taylor denominator,
we can concatenate the numerator 
\begin{align}
	\mu^4 b_4^{[a,b,c,f]} &= {1\over \la 12...61\ra} 
	\sum_{l_\pm}\left[  
	{\mu^4\la \chi |l_cbc(\mi l_c)q_{ce}efa|b\ra
	\over (s_{\mi l_cc}-\mu^2)\la b\chi\ra 
	(s_{el_f}-\mu^2)}\right]\biggr|_{\mathcal{O}(\mu^6)}\; .
\end{align}
By noting that
\begin{align}
	s_{\mi l_cc}-\mu^2 = -2 l_c \cdot p_c \; ,
	\notag \\
	2l_c\cdot p_b = l_b^2-\mu^2= 0 \; ,
\end{align}
we can commute the $l_c$ through up to $l_c\cdot l_c=\mu^2$ 
\begin{align}
	\mu^4 b_4^{[a,b,c,f]} &= {1\over \la 12 ...61\ra} 
	\sum_{l_\pm}\left[  
	{\mu^4[b|(\mi l_c)q_{ce}efa|b\ra
	\over 
	(s_{el_f}-\mu^2)}+\mathcal{O}(\mu^6)\right]\biggr|_{\mathcal{O}(\mu^6)}\; , 
\end{align}
and upon noting that by commuting $l_c$ through all
the other terms in $[b|\cdots |b\ra=\tr_+(b \cdots )$
\begin{align}
	[b|(\mi l_c)q_{ce}efa|b\ra
			&= -2l_c\cdot q_{ce}[b|afa|b\ra- 
			[b|(\mi l_c)q_{ce}efa|b\ra
		\notag \\
		& =  {1\over 2} (s_{el_f}-\mu^2)s_{fa}s_{ab}\,,
\end{align}
the one-mass box coefficient falls out without explicitly 
entering the solutions $l_\pm$, as
\begin{align}
	b_4^{[a,b,c,f]} &= 
	-{1\over \la 12 ...61\ra} 
	s_{fa}s_{ab} = {1\over 2}\tr(fabq_{cf}) \quad .
\end{align}
The two-mass box coefficients can be simplified in a similar way,
but to diversify the demonstration 
we show how to manipulate the expression 
with the $\alpha$ form of solution from equation~\bref{eq:all}.
\begin{align}
	\mu^4 b_4^{[a,c,d,f]} =& \sum_{l_\pm}\left[
		{\mu^2 [ab]\over \la ab \ra (s_{-l_a a}-\mu^2) } 
	\times 
	{\la \chi | l_c|c]\over \la c\chi \ra}
\times
	{\mu^2 [de]\over \la de \ra (s_{-l_d d}-\mu^2) }
	\times
	{\la \chi | l_f|f]\over \la f\chi \ra}
	\right]\biggr|_{\mathcal{O}(\mu^6)}
	\notag \\
	=& 
		{\mu^4 [ab][de]\over \la ab \ra \la de \ra} 
	\biggl[
	{
	\alpha [c|fq_{ac}|c]
	\alpha [f|cq_{df}|f]
	\over 4s^2_{af}(2(\alpha \bar{l}_{d+}+(1-\alpha)\bar{l}_{d-})
	\cdot p_d)
	(2(\alpha \bar{l}_{a+}+(1-\alpha)\bar{l}_{a-})\cdot p_a)}
+
	\notag \\
	&
	\hspace{0.8in}{
		(1-\alpha) [c|fq_{ac}|c]
	(1-\alpha) [f|cq_{df}|f]
	\over 4s^2_{af}(2((1-\alpha )\bar{l}_{d+}+
	\alpha\bar{l}_{d-})
	\cdot p_d)
	(2((1-\alpha) 
	\bar{l}_{a+}+\alpha\bar{l}_{a-})\cdot p_a)}
	\biggr]\biggr|_{\mathcal{O}(\mu^6)}\; ;
\end{align}
we can simply take the $\mu^2\rightarrow 0$ limit for the
content of the bracket
which implies, if we choose $\alpha = \alpha_+$,
that $\alpha\rightarrow 1$
\begin{align}
	\mu^4b_4^{[a,c,d,f]} 
	=& 
		{\mu^4 [ab][de]\over \la ab \ra \la de \ra} 
	\biggl[
	{
	[c|fq_{ac}|c]
	[f|cq_{df}|f]
	\over 4s^2_{af}(2\bar{l}_{d+}\cdot p_d
	2\bar{l}_{a+}\cdot p_a)
	}
	+\mathcal{O}(\mu^2)\biggr]\biggr|_{\mathcal{O}(\mu^6)}
\notag \\
	=& 
		{\mu^4 [ab][de]\over \la ab \ra \la de \ra} 
	\biggl[
	{
	[c|fq_{ac}|c]
	[f|cq_{df}|f]
	\over \la f|q_{df}dc|f]\la c|q_{ac}af|c]}
	+\mathcal{O}(\mu^2)\biggr]\biggr|_{\mathcal{O}(\mu^6)}
\notag \\
	=& 
		{\mu^4 \over \la 12...61\ra } 
	\biggl[
	\la f|q_{ac}|c]
	\la c|q_{df}|f]
	+\mathcal{O}(\mu^2)\biggr]\biggr|_{\mathcal{O}(\mu^6)};
\end{align}
thus, the coefficient is
\begin{align}
b_4^{[a,c,d,f]} 
	=&{1\over 2} {\tr(cq_{ac}fq_{df})\over \la 12 ... 61\ra }
	\quad .
\end{align}
The penta-cut in equation~\bref{eq:b5} 
can be used to compute the pentagon coefficient
\begin{align}
	b_5^{[a,b,c,d,e]} = c_0^3{[a|l_a|\chi\ra\over \la a \chi \ra}
	\times {[b|l_b|\chi\ra\over \la b \chi \ra}
	\times {[c|l_c|\chi\ra\over \la c \chi \ra}
	\times {[d|l_d|\chi\ra\over \la d \chi \ra}
	\times
	{\mu^2[ef]\over \la ef \ra (s_{\mi l_e e}-\mu^2)} \; ;
\end{align}
using the adjacent-three-point identity~\bref{eq:3t4} 
substituting in the solution~\bref{eq:5cutsol} gives
\begin{align}
	b_5^{[a,b,c,d,e]} &= {[ab]\over \la a b\ra}
	\times {[cd]\over \la c d\ra}
	\times
	{[ef]\over \la ef \ra }\times {1\over 2}\xi_{f}\; ;
	\notag \\
	&=
	{1\over \la 12...61\ra} 
	{\tr_+(abcdef)}{1\over 2}\xi_{f}
	\notag \\
	&=
	{1\over \la 12...61\ra} 
	\left( {\tr(123456)}{1\over 2}\xi_{f} - \tr_5(abcd)
	\right)
\end{align}
where upon summing over cycles the hexagon function can be
reconstructed from the first term in the final line.

Putting all the coefficients together reconstructs the 
full amplitude 
\begin{align}
	A_6^{\rm AP}=&
	{1\over \la 12...61\ra} 
\biggl[
		-\sum_{1<j_1<j_2\leq n}^n
	\tr\bigl(j_1q_{j_1+1,j_2-1}j_2q_{j_2+1,j_1-1}\bigr)
	I_4^{(j_1,j_2)}[\mu^4]
	\notag \\
	&	+ \sum_{j=1}^n\tr_5(j+1,j+2,j+3,j+4)
	I^{(j)}_5[\mu^6]
	+\tr(123456)I_6[\mu^6 ]\biggr] \; ,
\end{align}
which matches equation~\bref{eq:a6} upon shifting of the integrals
$I[\mu^{2r}]$ using equation~\bref{eq:mue}.

\subsection{Reduction to finite limit}
To verify our results we can check that taking the limit $\epsilon\rightarrow 0$
reproduces known results.
This is a strong check, as delicate cancellations need to occur between integral 
coefficients which do not follow obviously from their functional form.

The AP amplitude, to all orders in epsilon, is
\begin{align}
	A_n^{\rm AP} =
	\sum_{\mathcal{P}^{2me}_4\subset \lbrace 1,..,n\rbrace}
	b^\mathcal{P}_4 I^\mathcal{P}_4[\mu^4]
	+
	\sum_{\mathcal{P}_5\subset \lbrace 1,..,n\rbrace}
	b^\mathcal{P}_5 I^\mathcal{P}_5[\mu^6]\; ,
	\label{eq:apae}
\end{align}
where $\mathcal{P}_4^{2me}$ consists of the ``two-mass-easy" ordered
subsets of $\lbrace 1,...,n\rbrace$ of the form
\begin{align}
	\lbrace i_1,i_3-1,i_3,i_1-1\rbrace \; ,
\end{align}
the subsets $\mathcal{P}_5$ are all five-element 
subsets of $\lbrace 1,...,n\rbrace$,
and the $b^\mathcal{P}_4$ and $b^\mathcal{P}_5$ are defined in 
equations~\bref{eq:b4} and~\bref{eq:b5} respectively.
The $\epsilon\rightarrow 0$ limit should yield
\begin{align}
	 A^{\rm AP}_n= \sum_{1\leq i_1<i_2<i_3<i_4\leq n}
	 {\tr_-(i_1i_2i_3i_4)\over \la 12...n1 \ra}
	 +\mathcal{O}(\epsilon)\quad .
	\label{eq:ape0}
\end{align}
Upon making the decomposition $\tr_- = {1\over 2}(
\tr - \tr_5 )$ and taking into account the limits of the integral
\begin{align}
	\epsilon(1-\epsilon)I_4 &= {1\over 6}+\mathcal{O}(\epsilon)
	\; ,
	\notag \\
	\epsilon(1-\epsilon)I_5 &= {1\over 24}+\mathcal{O}(\epsilon)
	\; ,
\end{align}
then as shown by BDDK~\cite{Bern:1996ja} 
\begin{align}
	\sum_{\mathcal{P}^{2me}_4\subset \lbrace 1,..,n\rbrace}
	b^\mathcal{P}_4 I^\mathcal{P}_4[\mu^4]
	=
	 \sum_{1\leq i_1<i_2<i_3<i_4\leq n}
	 {\tr(i_1i_2i_3i_4)\over 2\la 12...n1 \ra} \; .
\end{align}
Equally it must hold true that 
\begin{align}
	\sum_{\mathcal{P}_5\subset \lbrace 1,..,n\rbrace}
	b^\mathcal{P}_5 I^\mathcal{P}_5[\mu^6] = 
	 -\sum_{1\leq i_1<i_2<i_3<i_4\leq n}
	 {\tr_5(i_1i_2i_3i_4)\over 2\la 12...n1 \ra} \quad  .
	 \label{eq:fl}
\end{align}
This does not appear to 
fall out naturally from the functional forms 
of the coefficients
$b^\mathcal{P}_5$, but as the solutions and functional form 
are completely determined, it provides a nontrivial check
on the validity of the expression~\bref{eq:b5}. 
We confirm that this holds numerically up to $n=17$.

\subsection{Transcendental structure}
\label{sec:transtr}
For $n=4,5$, and before performing the dimension shift, 
the MHV amplitude is  a pure function in the sense of \cite{ArkaniHamed:2010gh}.
That is to say, taking the $n=5$ case in particular, 
it
 can be written in the form
\begin{align}
	A_5^{\mathcal{N}=4}\left(1^-,2^-,3^+,4^+,5^+ \right)
	= A_5^{\rm tree}(1^-,2^-,3^+,4^+,5^+)f_5^{{\rm MPL}}
	(\lbrace {s_{i,i+1}}\rbrace , \epsilon) \; ,
\end{align}
where $f_5^{\rm MPL} $ is a linear combination of multiple 
polylogarithms without any additional kinematic coefficients.
This can be seen
by noticing that the coefficients appearing inside the brackets in  equation~\bref{eq:a5}
precisely cancel the leading singularities of the corresponding integrals,
leaving the normalized versions of  $I^{4-2\epsilon}_4,I^{6-2\epsilon}_5$, which  belong to the  uniformly transcendental basis 
used in~\cite{Abreu:2014cla}.

To see whether the pattern holds at $n=6$, we must check the leading singularity of the hexagon integral in equation~\bref{eq:a6}.
The amplitude before the dimension shift is
\begin{align}
	A^{\mathcal{N}=4}_6 =& {\la ij \ra^4\over
	4\la 12...61 \ra}\biggl[
		-\sum_{j_1,j_2=1}^n
		\tr\bigl(j_1q_{j_1+1,j_2-1}j_2
		q_{j_2+1,j_1-1}\bigr)I_4^{4-2\epsilon;(j_1,j_2)} 
	\notag \\
	&	-2\epsilon\left( \sum_{j=1}^n\tr_5(j+1,j+2,j+3,j+4)
	I^{6-2\epsilon,(j)}_5
	+\tr(123456)I^{6-2\epsilon}_6\right)\biggr]\quad .
\label{eq:n46}
\end{align}
Similarly to the five-point case, the boxes  and pentagon integrals 
are multiplied by the coefficients
cancelling their leading singularities, 
$\tr\bigl(j_1q_{j_1+1,j_2-1}j_2
q_{j_2+1,j_1-1}\bigr)$ 
and $\tr_5(j+1,j+2,j+3,j+4)$ 
respectively.\footnote{See~\cite{Jehu:2020xip} for 
representations of these reduction coefficients as determinants.}
The hexagon breaks down
entirely to 
pentagons~\cite{Bern:1993kr,Jehu:2020xip}
\begin{align}
	I_6  = {1\over 2}\sum_{j=1}^6 \xi_j I_5^{(j)}
\end{align}
where 
\begin{align}
	\xi_j= 2{(-1)^{j}\tr_5(j+1,j+2,j+3,j+4)\over
			\tr_5(123456)}
\end{align}
and thus the purity is broken by terms with the
rational prefactor
\begin{align}
(-1)^{j}{\tr(123456)\over
			\tr_5(123456)} \; ,
\end{align}
where the denominator is
the leading 
singularity\footnote{As the hexagon breaks down 
entirely to pentagons, the notion of it having a leading singularity
could be considered dubious. Here we use the same 
definition as in~\cite{Abreu:2014cla}, with a natural representation as a kinematic determinant} of the hexagon.
It should also be highlighted that the dimension shifted hexagon
does not contribute to the finite rational ($\epsilon^0$) piece of the 
AP amplitude as
\begin{align}
	\epsilon(1-\epsilon)I_6^{8-2\epsilon}= \mathcal{O}(\epsilon)\quad .
\end{align}

Thus while the 
 MHV amplitude fails to preserve purity at $n=6$,
 it is still possible  to match the pure structure 
to the finite contribution of the AP amplitude. 
However, this matching becomes more complicated for $n \geq 7$. 

The fact that MHV integrands in $\mathcal{N}=4$ SYM have dlog forms \cite{CaronHuot:2011ky,Arkani-Hamed:2016byb} led to the hope that the integrated expressions would be pure functions.
Of course, the dlog forms for MHV amplitudes cannot be integrated directly in four dimensions, but dimensional regularisation is not a priori incompatible with a similar analysis of transcendental behaviour. For example, 
it was mentioned in  \cite{Henn:2014qga} that dlog forms can naturally accommodate additional factors needed in dimensional regularisation, leading to results of uniform transcendental weight. 
 We have demonstrated by our explicit reduction that they are indeed of uniform transcendental weight in $D=4-2\epsilon$ dimensions for all $n$, but already the case $n=6$ shows that they do not satisfy the stronger condition of purity.

\newpage

%% file: conclusion-sec.tex

\section{Conclusion and Outlook}
In this article we have proved the dimension-shift 
conjecture in equation~\bref{eq:conj} 
by using unitarity cuts with a massive spinor-helicity formalism. 
We have also demonstrated how to compute 
the one-loop amplitudes in the conjecture to all orders in epsilon and 
at all multiplicities.
This is facilitated by the simplicity of the
pentagon cut solution~\bref{eq:5cutsol}.

Although one-loop amplitudes have for the most part long
been computable to any extent required by phenomenological 
applications, these results demonstrate that 
these two simple classes of
amplitude can be understood at a level of completeness not
previously realised.
Extending these results to other helicity configurations
is also achievable, by including dimension-shifted
bubble and triangle cuts and
using a limiting procedure to resolve the extra 
unfixed parameters in the cut 
constraint~\cite{Forde:2007mi,Badger:2008cm}.
For any given configuration, one would also need the appropriate lower-point tree amplitudes
with two massive legs.

Applying similar techniques to multi-loop amplitudes requires 
a general understanding of how to fix a basis as was done
in section~\ref{sec:basis}.
We hope the one-loop techniques consolidated and developed 
in this work can provide insight and guidance into
pushing understanding of multi-loop amplitudes 
beyond their current level.

Moreover, the theorem~\bref{eq:conj} and the proof given here
bridges $\mathcal{N}=4$ super-Yang-Mills, with its
remarkable simplicity, and more realistic theories like QCD.
We hope that further simplifications and unifications of this
kind follow down the line.

%% file: conj.bbl
\begin{thebibliography}{10}
\providecommand{\href}[2]{#2}
\providecommand{\arxivref}[2]{\href{http://arxiv.org/abs/#1}{#2}}
\providecommand{\doiref}[2]{\href{http://dx.doi.org/#1}{#2}}
\providecommand{\nbbstauthor}[1]{#1}
\providecommand{\nbbstjournal}[1]{\textsf{#1}}
\providecommand{\nbbsttitle}[1]{\textit{#1}}
\providecommand{\nbbsturl}[1]{\texttt{#1}}
\providecommand{\nbbsteprint}[1]{\texttt{#1}}
\providecommand{\nbbststyle}{\raggedright\small\parskip0pt}
\nbbststyle

\bibitem{Caron-Huot:2019bsq}
\nbbstauthor{S.~Caron-Huot, L.~J.~Dixon, F.~Dulat, M.~Von~Hippel, A.~J.~McLeod
  and G.~Papathanasiou},
\nbbsttitle{``{The Cosmic Galois Group and Extended Steinmann Relations for
  Planar $\mathcal{N} = 4$ SYM Amplitudes}''},
\nbbstjournal{\doiref{10.1007/JHEP09(2019)061}{JHEP~1909,~061~(2019)}},
\nbbsteprint{\arxivref{1906.07116}{arxiv:1906.07116}}.

\bibitem{Dixon:2020cnr}
\nbbstauthor{L.~J.~Dixon and Y.-T.~Liu},
\nbbsttitle{``{Lifting Heptagon Symbols to Functions}''},
\nbbsteprint{\arxivref{2007.12966}{arxiv:2007.12966}}.

\bibitem{Elvang:2015rqa}
\nbbstauthor{H.~Elvang and Y.-t.~Huang},
\nbbsttitle{``{Scattering Amplitudes in Gauge Theory and Gravity}''},
Cambridge University Press (2015).

\bibitem{Arkani-Hamed:2016byb}
\nbbstauthor{N.~Arkani-Hamed, J.~L.~Bourjaily, F.~Cachazo, A.~B.~Goncharov,
  A.~Postnikov and J.~Trnka},
\nbbsttitle{``{Grassmannian Geometry of Scattering Amplitudes}''},
Cambridge University Press (2016).

\bibitem{Bern:1994ju}
\nbbstauthor{Z.~Bern, L.~J.~Dixon, D.~C.~Dunbar and D.~A.~Kosower},
\nbbsttitle{``{One loop gauge theory amplitudes with an arbitrary number of
  external legs}''},
\nbbsteprint{\arxivref{hep-ph/9405248}{hep-ph/9405248}},
in: \nbbsttitle{``{Workshop on Continuous Advances in QCD Minneapolis,
  Minnesota, February 18-20, 1994}''},
pp.~3-21.

\bibitem{Bern:1994cg}
\nbbstauthor{Z.~Bern, L.~J.~Dixon, D.~C.~Dunbar and D.~A.~Kosower},
\nbbsttitle{``{Fusing gauge theory tree amplitudes into loop amplitudes}''},
\nbbstjournal{\doiref{10.1016/0550-3213(94)00488-Z}{Nucl.~Phys.~B435,~59~(1995)}},
\nbbsteprint{\arxivref{hep-ph/9409265}{hep-ph/9409265}}.

\bibitem{Britto:2004nc}
\nbbstauthor{R.~Britto, F.~Cachazo and B.~Feng},
\nbbsttitle{``{Generalized unitarity and one-loop amplitudes in N=4
  super-Yang-Mills}''},
\nbbstjournal{\doiref{10.1016/j.nuclphysb.2005.07.014}{Nucl.~Phys.~B725,~275~(2005)}},
\nbbsteprint{\arxivref{hep-th/0412103}{hep-th/0412103}}.

\bibitem{CaronHuot:2012ab}
\nbbstauthor{S.~Caron-Huot and K.~J.~Larsen},
\nbbsttitle{``{Uniqueness of two-loop master contours}''},
\nbbstjournal{\doiref{10.1007/JHEP10(2012)026}{JHEP~1210,~026~(2012)}},
\nbbsteprint{\arxivref{1205.0801}{arxiv:1205.0801}}.

\bibitem{Badger:2013gxa}
\nbbstauthor{S.~Badger, H.~Frellesvig and Y.~Zhang},
\nbbsttitle{``{A Two-Loop Five-Gluon Helicity Amplitude in QCD}''},
\nbbstjournal{\doiref{10.1007/JHEP12(2013)045}{JHEP~1312,~045~(2013)}},
\nbbsteprint{\arxivref{1310.1051}{arxiv:1310.1051}}.

\bibitem{Badger:2017jhb}
\nbbstauthor{S.~Badger, C.~Brønnum-Hansen, H.~B.~Hartanto and T.~Peraro},
\nbbsttitle{``{First look at two-loop five-gluon scattering in QCD}''},
\nbbstjournal{\doiref{10.1103/PhysRevLett.120.092001}{Phys.~Rev.~Lett.~120,~092001~(2018)}},
\nbbsteprint{\arxivref{1712.02229}{arxiv:1712.02229}}.

\bibitem{Abreu:2017hqn}
\nbbstauthor{S.~Abreu, F.~Febres~Cordero, H.~Ita, B.~Page and M.~Zeng},
\nbbsttitle{``{Planar Two-Loop Five-Gluon Amplitudes from Numerical
  Unitarity}''},
\nbbstjournal{\doiref{10.1103/PhysRevD.97.116014}{Phys.~Rev.~D97,~116014~(2018)}},
\nbbsteprint{\arxivref{1712.03946}{arxiv:1712.03946}}.

\bibitem{Peraro:2016wsq}
\nbbstauthor{T.~Peraro},
\nbbsttitle{``{Scattering amplitudes over finite fields and multivariate
  functional reconstruction}''},
\nbbstjournal{\doiref{10.1007/JHEP12(2016)030}{JHEP~1612,~030~(2016)}},
\nbbsteprint{\arxivref{1608.01902}{arxiv:1608.01902}}.

\bibitem{Badger:2018enw}
\nbbstauthor{S.~Badger, C.~Brønnum-Hansen, H.~B.~Hartanto and T.~Peraro},
\nbbsttitle{``{Analytic helicity amplitudes for two-loop five-gluon scattering:
  the single-minus case}''},
\nbbstjournal{\doiref{10.1007/JHEP01(2019)186}{JHEP~1901,~186~(2019)}},
\nbbsteprint{\arxivref{1811.11699}{arxiv:1811.11699}}.

\bibitem{Abreu:2018zmy}
\nbbstauthor{S.~Abreu, J.~Dormans, F.~Febres~Cordero, H.~Ita and B.~Page},
\nbbsttitle{``{Analytic Form of Planar Two-Loop Five-Gluon Scattering
  Amplitudes in QCD}''},
\nbbstjournal{\doiref{10.1103/PhysRevLett.122.082002}{Phys.~Rev.~Lett.~122,~082002~(2019)}},
\nbbsteprint{\arxivref{1812.04586}{arxiv:1812.04586}}.

\bibitem{Abreu:2019odu}
\nbbstauthor{S.~Abreu, J.~Dormans, F.~Febres~Cordero, H.~Ita, B.~Page and
  V.~Sotnikov},
\nbbsttitle{``{Analytic Form of the Planar Two-Loop Five-Parton Scattering
  Amplitudes in QCD}''},
\nbbstjournal{\doiref{10.1007/JHEP05(2019)084}{JHEP~1905,~084~(2019)}},
\nbbsteprint{\arxivref{1904.00945}{arxiv:1904.00945}}.

\bibitem{Gehrmann:2015bfy}
\nbbstauthor{T.~Gehrmann, J.~M.~Henn and N.~A.~Lo~Presti},
\nbbsttitle{``{Analytic form of the two-loop planar five-gluon
  all-plus-helicity amplitude in QCD}''},
\nbbstjournal{\doiref{10.1103/PhysRevLett.116.189903,
  10.1103/PhysRevLett.116.062001}{Phys.~Rev.~Lett.~116,~062001~(2016)}},
\nbbsteprint{\arxivref{1511.05409}{arxiv:1511.05409}},
[Erratum: Phys. Rev. Lett.116,no.18,189903(2016)].

\bibitem{Dunbar:2016aux}
\nbbstauthor{D.~C.~Dunbar and W.~B.~Perkins},
\nbbsttitle{``{Two-loop five-point all plus helicity Yang-Mills amplitude}''},
\nbbstjournal{\doiref{10.1103/PhysRevD.93.085029}{Phys.~Rev.~D93,~085029~(2016)}},
\nbbsteprint{\arxivref{1603.07514}{arxiv:1603.07514}}.

\bibitem{Dunbar:2017nfy}
\nbbstauthor{D.~C.~Dunbar, J.~H.~Godwin, G.~R.~Jehu and W.~B.~Perkins},
\nbbsttitle{``{Analytic all-plus-helicity gluon amplitudes in QCD}''},
\nbbstjournal{\doiref{10.1103/PhysRevD.96.116013}{Phys.~Rev.~D96,~116013~(2017)}},
\nbbsteprint{\arxivref{1710.10071}{arxiv:1710.10071}}.

\bibitem{Badger:2019djh}
\nbbstauthor{S.~Badger, D.~Chicherin, T.~Gehrmann, G.~Heinrich, J.~M.~Henn,
  T.~Peraro, P.~Wasser, Y.~Zhang and S.~Zoia},
\nbbsttitle{``{Analytic form of the full two-loop five-gluon all-plus helicity
  amplitude}''},
\nbbstjournal{\doiref{10.1103/PhysRevLett.123.071601}{Phys.~Rev.~Lett.~123,~071601~(2019)}},
\nbbsteprint{\arxivref{1905.03733}{arxiv:1905.03733}}.

\bibitem{Dunbar:2019fcq}
\nbbstauthor{D.~C.~Dunbar, J.~H.~Godwin, W.~B.~Perkins and J.~M.~Strong},
\nbbsttitle{``{Color Dressed Unitarity and Recursion for Yang-Mills Two-Loop
  All-Plus Amplitudes}''},
\nbbstjournal{\doiref{10.1103/PhysRevD.101.016009}{Phys.~Rev.~D~101,~016009~(2020)}},
\nbbsteprint{\arxivref{1911.06547}{arxiv:1911.06547}}.

\bibitem{Henn:2019mvc}
\nbbstauthor{J.~Henn, B.~Power and S.~Zoia},
\nbbsttitle{``{Conformal Invariance of the One-Loop All-Plus Helicity
  Scattering Amplitudes}''},
\nbbstjournal{\doiref{10.1007/JHEP02(2020)019}{JHEP~2002,~019~(2020)}},
\nbbsteprint{\arxivref{1911.12142}{arxiv:1911.12142}}.

\bibitem{Dunbar:2020wdh}
\nbbstauthor{D.~C.~Dunbar, W.~B.~Perkins and J.~M.~Strong},
\nbbsttitle{``{$n$-point QCD two-loop amplitude}''},
\nbbstjournal{\doiref{10.1103/PhysRevD.101.076001}{Phys.~Rev.~D~101,~076001~(2020)}},
\nbbsteprint{\arxivref{2001.11347}{arxiv:2001.11347}}.

\bibitem{Bern:1996ja}
\nbbstauthor{Z.~Bern, L.~J.~Dixon, D.~C.~Dunbar and D.~A.~Kosower},
\nbbsttitle{``{One loop selfdual and N=4 superYang-Mills}''},
\nbbstjournal{\doiref{10.1016/S0370-2693(96)01676-0}{Phys.~Lett.~B394,~105~(1997)}},
\nbbsteprint{\arxivref{hep-th/9611127}{hep-th/9611127}}.

\bibitem{Cangemi:1996pf}
\nbbstauthor{D.~Cangemi},
\nbbsttitle{``{Selfduality and maximally helicity violating QCD amplitudes}''},
\nbbstjournal{\doiref{10.1142/S0217751X97000943}{Int.~J.~Mod.~Phys.~A~12,~1215~(1997)}},
\nbbsteprint{\arxivref{hep-th/9610021}{hep-th/9610021}}.

\bibitem{Chalmers:1996rq}
\nbbstauthor{G.~Chalmers and W.~Siegel},
\nbbsttitle{``{The Selfdual sector of QCD amplitudes}''},
\nbbstjournal{\doiref{10.1103/PhysRevD.54.7628}{Phys.~Rev.~D54,~7628~(1996)}},
\nbbsteprint{\arxivref{hep-th/9606061}{hep-th/9606061}}.

\bibitem{Bardeen:1995gk}
\nbbstauthor{W.~A.~Bardeen},
\nbbsttitle{``{Selfdual Yang-Mills theory, integrability and multiparton
  amplitudes}''},
\nbbstjournal{\doiref{10.1143/PTPS.123.1}{Prog.~Theor.~Phys.~Suppl.~123,~1~(1996)}}.

\bibitem{Chattopadhyay2020}
\nbbstauthor{P.~Chattopadhyay and K.~Krasnov},
\nbbsttitle{``One-loop same helicity four-point amplitude from shifts''},
\nbbsteprint{\arxivref{2002.11390}{arxiv:2002.11390}}.

\bibitem{Popov:1998pc}
\nbbstauthor{A.~Popov},
\nbbsttitle{``{Selfdual Yang-Mills: Symmetries and moduli space}''},
\nbbstjournal{\doiref{10.1142/S0129055X99000350}{Rev.~Math.~Phys.~11,~1091~(1999)}},
\nbbsteprint{\arxivref{hep-th/9803183}{hep-th/9803183}}.

\bibitem{Ooguri1995}
\nbbstauthor{H.~Ooguri and C.~Vafa},
\nbbsttitle{``All Loop N=2 String Amplitudes''},
\nbbsteprint{\arxivref{hep-th/9505183}{hep-th/9505183}}.

\bibitem{Green:1982sw}
\nbbstauthor{M.~B.~Green, J.~H.~Schwarz and L.~Brink},
\nbbsttitle{``{N=4 Yang-Mills and N=8 Supergravity as Limits of String
  Theories}''},
\nbbstjournal{\doiref{10.1016/0550-3213(82)90336-4}{Nucl.~Phys.~B~198,~474~(1982)}}.

\bibitem{Bern:1991aq}
\nbbstauthor{Z.~Bern and D.~A.~Kosower},
\nbbsttitle{``{The Computation of loop amplitudes in gauge theories}''},
\nbbstjournal{\doiref{10.1016/0550-3213(92)90134-W}{Nucl.~Phys.~B379,~451~(1992)}}.

\bibitem{Bern:1991an}
\nbbstauthor{Z.~Bern and D.~C.~Dunbar},
\nbbsttitle{``{A Mapping between Feynman and string motivated one loop rules in
  gauge theories}''},
\nbbstjournal{\doiref{10.1016/0550-3213(92)90135-X}{Nucl.~Phys.~B379,~562~(1992)}}.

\bibitem{Ahmadiniaz2020}
\nbbstauthor{N.~Ahmadiniaz and C.~Schubert},
\nbbsttitle{``Off-shell Ward identities for N-gluon amplitudes''},
\nbbsteprint{\arxivref{2001.00885}{arxiv:2001.00885}}.

\bibitem{Ahmadiniaz2020a}
\nbbstauthor{N.~Ahmadiniaz, V.~M.~B.~Guzman, F.~Bastianelli, O.~Corradini,
  J.~P.~Edwards and C.~Schubert},
\nbbsttitle{``Worldline master formulas for the dressed electron propagator,
  part 1: Off-shell amplitudes''},
\nbbsteprint{\arxivref{2004.01391}{arxiv:2004.01391}}.

\bibitem{Bonocore2021}
\nbbstauthor{D.~Bonocore},
\nbbsttitle{``Asymptotic dynamics on the worldline for spinning particles''},
\nbbstjournal{\doiref{10.1007/JHEP02(2021)007}{JHEP~2102,~007~(2021)}},
\nbbsteprint{\arxivref{2009.07863}{arxiv:2009.07863}}.

\bibitem{Mogull2020}
\nbbstauthor{G.~Mogull, J.~Plefka and J.~Steinhoff},
\nbbsttitle{``Classical black hole scattering from a worldline quantum field
  theory''},
\nbbsteprint{\arxivref{2010.02865}{arxiv:2010.02865}}.

\bibitem{Schabinger2011}
\nbbstauthor{R.~M.~Schabinger},
\nbbsttitle{``One-Loop N = 4 Super Yang-Mills Scattering Amplitudes to All
  Orders in the Dimensional Regularization Parameter''},
\nbbsteprint{\arxivref{1103.2769}{arxiv:1103.2769}}.

\bibitem{Stieberger2006}
\nbbstauthor{S.~Stieberger and T.~R.~Taylor},
\nbbsttitle{``Amplitude for N-Gluon Superstring Scattering''},
\nbbsteprint{\arxivref{hep-th/0607184}{hep-th/0607184}}.

\bibitem{Grisaru:1977px}
\nbbstauthor{M.~T.~Grisaru and H.~N.~Pendleton},
\nbbsttitle{``{Some Properties of Scattering Amplitudes in Supersymmetric
  Theories}''},
\nbbstjournal{\doiref{10.1016/0550-3213(77)90277-2}{Nucl.~Phys.~B124,~81~(1977)}}.

\bibitem{Dixon1996}
\nbbstauthor{L.~Dixon},
\nbbsttitle{``Calculating Scattering Amplitudes Efficiently''},
\nbbsteprint{\arxivref{hep-ph/9601359}{hep-ph/9601359}}.

\bibitem{Bern1992a}
\nbbstauthor{Z.~Bern},
\nbbsttitle{``{String based perturbative methods for gauge theories}''},
\nbbsteprint{\arxivref{hep-ph/9304249}{hep-ph/9304249}},
in: \nbbsttitle{``{Proceedings, Theoretical Advanced Study Institute (TASI 92):
  From Black Holes and Strings to Particles: Boulder, USA, June 1-26, 1992}''},
pp.~0471-536.

\bibitem{Badger:2008cm}
\nbbstauthor{S.~D.~Badger},
\nbbsttitle{``{Direct Extraction Of One Loop Rational Terms}''},
\nbbstjournal{\doiref{10.1088/1126-6708/2009/01/049}{JHEP~0901,~049~(2009)}},
\nbbsteprint{\arxivref{0806.4600}{arxiv:0806.4600}}.

\bibitem{Bern:1993kr}
\nbbstauthor{Z.~Bern, L.~J.~Dixon and D.~A.~Kosower},
\nbbsttitle{``{Dimensionally regulated pentagon integrals}''},
\nbbstjournal{\doiref{10.1016/0550-3213(94)90398-0}{Nucl.~Phys.~B412,~751~(1994)}},
\nbbsteprint{\arxivref{hep-ph/9306240}{hep-ph/9306240}}.

\bibitem{Jehu:2020xip}
\nbbstauthor{G.~R.~Jehu},
\nbbsttitle{``{Symmetric reduction of high-multiplicity one-loop integrals and
  maximal cuts}''},
\nbbsteprint{\arxivref{2010.16266}{arxiv:2010.16266}}.

\bibitem{Abreu:2017ptx}
\nbbstauthor{S.~Abreu, R.~Britto, C.~Duhr and E.~Gardi},
\nbbsttitle{``{Cuts from residues: the one-loop case}''},
\nbbstjournal{\doiref{10.1007/JHEP06(2017)114}{JHEP~1706,~114~(2017)}},
\nbbsteprint{\arxivref{1702.03163}{arxiv:1702.03163}}.

\bibitem{Bianchi:2008pu}
\nbbstauthor{M.~Bianchi, H.~Elvang and D.~Z.~Freedman},
\nbbsttitle{``{Generating Tree Amplitudes in N=4 SYM and N = 8 SG}''},
\nbbstjournal{\doiref{10.1088/1126-6708/2008/09/063}{JHEP~0809,~063~(2008)}},
\nbbsteprint{\arxivref{0805.0757}{arxiv:0805.0757}}.

\bibitem{Mahlon:1993si}
\nbbstauthor{G.~Mahlon},
\nbbsttitle{``{Multi - gluon helicity amplitudes involving a quark loop}''},
\nbbstjournal{\doiref{10.1103/PhysRevD.49.4438}{Phys.~Rev.~D49,~4438~(1994)}},
\nbbsteprint{\arxivref{hep-ph/9312276}{hep-ph/9312276}}.

\bibitem{Bern:1993qk}
\nbbstauthor{Z.~Bern, G.~Chalmers, L.~J.~Dixon and D.~A.~Kosower},
\nbbsttitle{``{One loop N gluon amplitudes with maximal helicity violation via
  collinear limits}''},
\nbbstjournal{\doiref{10.1103/PhysRevLett.72.2134}{Phys.~Rev.~Lett.~72,~2134~(1994)}},
\nbbsteprint{\arxivref{hep-ph/9312333}{hep-ph/9312333}}.

\bibitem{Bern:1995db}
\nbbstauthor{Z.~Bern and A.~G.~Morgan},
\nbbsttitle{``{Massive loop amplitudes from unitarity}''},
\nbbstjournal{\doiref{10.1016/0550-3213(96)00078-8}{Nucl.~Phys.~B467,~479~(1996)}},
\nbbsteprint{\arxivref{hep-ph/9511336}{hep-ph/9511336}}.

\bibitem{Dittmaier:1998nn}
\nbbstauthor{S.~Dittmaier},
\nbbsttitle{``Weyl-van-der-Waerden formalism for helicity amplitudes of massive
  particles''},
\nbbsteprint{\arxivref{hep-ph/9805445}{hep-ph/9805445}}.

\bibitem{Arkani-Hamed:2017jhn}
\nbbstauthor{N.~Arkani-Hamed, T.-C.~Huang and Y.-t.~Huang},
\nbbsttitle{``{Scattering Amplitudes For All Masses and Spins}''},
\nbbsteprint{\arxivref{1709.04891}{arxiv:1709.04891}}.

\bibitem{Boels:2011zz}
\nbbstauthor{R.~H.~Boels and C.~Schwinn},
\nbbsttitle{``On-shell supersymmetry for massive multiplets''},
\nbbsteprint{\arxivref{1104.2280}{arxiv:1104.2280}}.

\bibitem{Craig:2011ws}
\nbbstauthor{N.~Craig, H.~Elvang, M.~Kiermaier and T.~Slatyer},
\nbbsttitle{``Massive amplitudes on the Coulomb branch of N=4 SYM''},
\nbbsteprint{\arxivref{1104.2050}{arxiv:1104.2050}}.

\bibitem{Kiermaier:2011cr}
\nbbstauthor{M.~Kiermaier},
\nbbsttitle{``The Coulomb-branch S-matrix from massless amplitudes''},
\nbbsteprint{\arxivref{1105.5385}{arxiv:1105.5385}}.

\bibitem{Elvang:2011ub}
\nbbstauthor{H.~Elvang, D.~Z.~Freedman and M.~Kiermaier},
\nbbsttitle{``Integrands for QCD rational terms and N=4 SYM from massive CSW
  rules''},
\nbbsteprint{\arxivref{1111.0635}{arxiv:1111.0635}}.

\bibitem{Eden1966}
\nbbstauthor{R.~J.~Eden, P.~V.~Landshoff, D.~I.~Olive and J.~C.~Polkinghorne},
\nbbsttitle{``{The analytic S-matrix}''},
Cambridge Univ. Press (1966),
Cambridge.

\bibitem{Bern:1994zx}
\nbbstauthor{Z.~Bern, L.~J.~Dixon, D.~C.~Dunbar and D.~A.~Kosower},
\nbbsttitle{``{One loop n point gauge theory amplitudes, unitarity and
  collinear limits}''},
\nbbstjournal{\doiref{10.1016/0550-3213(94)90179-1}{Nucl.~Phys.~B425,~217~(1994)}},
\nbbsteprint{\arxivref{hep-ph/9403226}{hep-ph/9403226}}.

\bibitem{Britto:2006sj}
\nbbstauthor{R.~Britto, B.~Feng and P.~Mastrolia},
\nbbsttitle{``{The Cut-constructible part of QCD amplitudes}''},
\nbbstjournal{\doiref{10.1103/PhysRevD.73.105004}{Phys.~Rev.~D73,~105004~(2006)}},
\nbbsteprint{\arxivref{hep-ph/0602178}{hep-ph/0602178}}.

\bibitem{Forde:2007mi}
\nbbstauthor{D.~Forde},
\nbbsttitle{``{Direct extraction of one-loop integral coefficients}''},
\nbbstjournal{\doiref{10.1103/PhysRevD.75.125019}{Phys.~Rev.~D75,~125019~(2007)}},
\nbbsteprint{\arxivref{0704.1835}{arxiv:0704.1835}}.

\bibitem{Dunbar:2009ax}
\nbbstauthor{D.~C.~Dunbar, W.~B.~Perkins and E.~Warrick},
\nbbsttitle{``{The Unitarity Method using a Canonical Basis Approach}''},
\nbbstjournal{\doiref{10.1088/1126-6708/2009/06/056}{JHEP~0906,~056~(2009)}},
\nbbsteprint{\arxivref{0903.1751}{arxiv:0903.1751}}.

\bibitem{Mastrolia:2009dr}
\nbbstauthor{P.~Mastrolia},
\nbbsttitle{``{Double-Cut of Scattering Amplitudes and Stokes' Theorem}''},
\nbbstjournal{\doiref{10.1016/j.physletb.2009.06.033}{Phys.~Lett.~B678,~246~(2009)}},
\nbbsteprint{\arxivref{0905.2909}{arxiv:0905.2909}}.

\bibitem{Bern:2005hs}
\nbbstauthor{Z.~Bern, L.~J.~Dixon and D.~A.~Kosower},
\nbbsttitle{``{On-shell recurrence relations for one-loop QCD amplitudes}''},
\nbbstjournal{\doiref{10.1103/PhysRevD.71.105013}{Phys.~Rev.~D71,~105013~(2005)}},
\nbbsteprint{\arxivref{hep-th/0501240}{hep-th/0501240}}.

\bibitem{Dunbar:2010wu}
\nbbstauthor{D.~C.~Dunbar, J.~H.~Ettle and W.~B.~Perkins},
\nbbsttitle{``{Augmented Recursion For One-loop Amplitudes}''},
\nbbstjournal{\doiref{10.1016/j.nuclphysbps.2010.08.022}{Nucl.~Phys.~Proc.~Suppl.~205-206,~74~(2010)}},
\nbbsteprint{\arxivref{1011.0559}{arxiv:1011.0559}},
in: \nbbsttitle{``{Proceedings, 10th DESY Workshop on Elementary Particle
  Theory: Loops and Legs in Quantum Field Theory: Woerlitz, Germany, April
  25-30, 2010}''},
pp.~74-79.

\bibitem{Alston:2012xd}
\nbbstauthor{S.~D.~Alston, D.~C.~Dunbar and W.~B.~Perkins},
\nbbsttitle{``{Complex Factorisation and Recursion for One-Loop Amplitudes}''},
\nbbstjournal{\doiref{10.1103/PhysRevD.86.085022}{Phys.~Rev.~D86,~085022~(2012)}},
\nbbsteprint{\arxivref{1208.0190}{arxiv:1208.0190}}.

\bibitem{Giele:2008ve}
\nbbstauthor{W.~T.~Giele, Z.~Kunszt and K.~Melnikov},
\nbbsttitle{``{Full one-loop amplitudes from tree amplitudes}''},
\nbbstjournal{\doiref{10.1088/1126-6708/2008/04/049}{JHEP~0804,~049~(2008)}},
\nbbsteprint{\arxivref{0801.2237}{arxiv:0801.2237}}.

\bibitem{Ellis:2008ir}
\nbbstauthor{R.~K.~Ellis, W.~T.~Giele, Z.~Kunszt and K.~Melnikov},
\nbbsttitle{``{Masses, fermions and generalized $D$-dimensional unitarity}''},
\nbbstjournal{\doiref{10.1016/j.nuclphysb.2009.07.023}{Nucl.~Phys.~B822,~270~(2009)}},
\nbbsteprint{\arxivref{0806.3467}{arxiv:0806.3467}}.

\bibitem{Ossola:2008xq}
\nbbstauthor{G.~Ossola, C.~G.~Papadopoulos and R.~Pittau},
\nbbsttitle{``{On the Rational Terms of the one-loop amplitudes}''},
\nbbstjournal{\doiref{10.1088/1126-6708/2008/05/004}{JHEP~0805,~004~(2008)}},
\nbbsteprint{\arxivref{0802.1876}{arxiv:0802.1876}}.

\bibitem{Cheung2009}
\nbbstauthor{C.~Cheung and D.~O'Connell},
\nbbsttitle{``{Amplitudes and Spinor-Helicity in Six Dimensions}''},
\nbbstjournal{\doiref{10.1088/1126-6708/2009/07/075}{JHEP~0907,~075~(2009)}},
\nbbsteprint{\arxivref{0902.0981}{arxiv:0902.0981}}.

\bibitem{Bern2010}
\nbbstauthor{Z.~Bern, J.~J.~Carrasco, T.~Dennen, Y.~tin~Huang and H.~Ita},
\nbbsttitle{``Generalized Unitarity and Six-Dimensional Helicity''},
\nbbsteprint{\arxivref{1010.0494}{arxiv:1010.0494}}.

\bibitem{Badger2017}
\nbbstauthor{S.~Badger, C.~Brønnum-Hansen, F.~Buciuni and D.~O'Connell},
\nbbsttitle{``{A unitarity compatible approach to one-loop amplitudes with
  massive fermions}''},
\nbbstjournal{\doiref{10.1007/JHEP06(2017)141}{JHEP~1706,~141~(2017)}},
\nbbsteprint{\arxivref{1703.05734}{arxiv:1703.05734}}.

\bibitem{Pauli:1949zm}
\nbbstauthor{W.~Pauli and F.~Villars},
\nbbsttitle{``{On the Invariant regularization in relativistic quantum
  theory}''},
\nbbstjournal{\doiref{10.1103/RevModPhys.21.434}{Rev.~Mod.~Phys.~21,~434~(1949)}}.

\bibitem{Forde:2005ue}
\nbbstauthor{D.~Forde and D.~A.~Kosower},
\nbbsttitle{``{All-multiplicity amplitudes with massive scalars}''},
\nbbstjournal{\doiref{10.1103/PhysRevD.73.065007}{Phys.~Rev.~D73,~065007~(2006)}},
\nbbsteprint{\arxivref{hep-th/0507292}{hep-th/0507292}}.

\bibitem{Rodrigo:2005eu}
\nbbstauthor{G.~Rodrigo},
\nbbsttitle{``Multigluonic scattering amplitudes of heavy quarks''},
\nbbsteprint{\arxivref{hep-ph/0508138}{hep-ph/0508138}}.

\bibitem{Ferrario:2006np}
\nbbstauthor{P.~Ferrario, G.~Rodrigo and P.~Talavera},
\nbbsttitle{``Compact multigluonic scattering amplitudes with heavy scalars and
  fermions''},
\nbbsteprint{\arxivref{hep-th/0602043}{hep-th/0602043}}.

\bibitem{Ochirov2018}
\nbbstauthor{A.~Ochirov},
\nbbsttitle{``{Helicity amplitudes for QCD with massive quarks}''},
\nbbstjournal{\doiref{10.1007/JHEP04(2018)089}{JHEP~1804,~089~(2018)}},
\nbbsteprint{\arxivref{1802.06730}{arxiv:1802.06730}}.

\bibitem{Schwinn:2006ca}
\nbbstauthor{C.~Schwinn and S.~Weinzierl},
\nbbsttitle{``{SUSY ward identities for multi-gluon helicity amplitudes with
  massive quarks}''},
\nbbstjournal{\doiref{10.1088/1126-6708/2006/03/030}{JHEP~0603,~030~(2006)}},
\nbbsteprint{\arxivref{hep-th/0602012}{hep-th/0602012}}.

\bibitem{Schwinn2005}
\nbbstauthor{C.~Schwinn and S.~Weinzierl},
\nbbsttitle{``{Scalar diagrammatic rules for Born amplitudes in QCD}''},
\nbbstjournal{\doiref{10.1088/1126-6708/2005/05/006}{JHEP~0505,~006~(2005)}},
\nbbsteprint{\arxivref{hep-th/0503015}{hep-th/0503015}}.

\bibitem{Passarino1979}
\nbbstauthor{G.~Passarino and M.~J.~G.~Veltman},
\nbbsttitle{``{One Loop Corrections for $e^+ e^-$ Annihilation Into $\mu^+
  \mu^-$ in the Weinberg Model}''},
\nbbstjournal{\doiref{10.1016/0550-3213(79)90234-7}{Nucl.~Phys.~B160,~151~(1979)}}.

\bibitem{Neerven1984}
\nbbstauthor{W.~L.~van~Neerven and J.~A.~M.~Vermaseren},
\nbbsttitle{``{Large loop integrals}''},
\nbbstjournal{\doiref{10.1016/0370-2693(84)90237-5}{Phys.~Lett.~137B,~241~(1984)}}.

\bibitem{Melrose1965}
\nbbstauthor{D.~B.~Melrose},
\nbbsttitle{``{Reduction of Feynman diagrams}''},
\nbbstjournal{\doiref{10.1007/BF02832919}{Nuovo~Cim.~40,~181~(1965)}}.

\bibitem{ArkaniHamed:2010gh}
\nbbstauthor{N.~Arkani-Hamed, J.~L.~Bourjaily, F.~Cachazo and J.~Trnka},
\nbbsttitle{``{Local Integrals for Planar Scattering Amplitudes}''},
\nbbstjournal{\doiref{10.1007/JHEP06(2012)125}{JHEP~1206,~125~(2012)}},
\nbbsteprint{\arxivref{1012.6032}{arxiv:1012.6032}}.

\bibitem{Abreu:2014cla}
\nbbstauthor{S.~Abreu, R.~Britto, C.~Duhr and E.~Gardi},
\nbbsttitle{``{From multiple unitarity cuts to the coproduct of Feynman
  integrals}''},
\nbbstjournal{\doiref{10.1007/JHEP10(2014)125}{JHEP~1410,~125~(2014)}},
\nbbsteprint{\arxivref{1401.3546}{arxiv:1401.3546}}.

\bibitem{CaronHuot:2011ky}
\nbbstauthor{S.~Caron-Huot},
\nbbsttitle{``{Superconformal symmetry and two-loop amplitudes in planar N=4
  super Yang-Mills}''},
\nbbstjournal{\doiref{10.1007/JHEP12(2011)066}{JHEP~1112,~066~(2011)}},
\nbbsteprint{\arxivref{1105.5606}{arxiv:1105.5606}}.

\bibitem{Henn:2014qga}
\nbbstauthor{J.~M.~Henn},
\nbbsttitle{``{Lectures on differential equations for Feynman integrals}''},
\nbbstjournal{\doiref{10.1088/1751-8113/48/15/153001}{J.~Phys.~A~48,~153001~(2015)}},
\nbbsteprint{\arxivref{1412.2296}{arxiv:1412.2296}}.

\end{thebibliography}
